\documentclass{vldb}
\usepackage{graphicx}
\usepackage{balance}  
\pdfoutput=1

\vldbTitle{A Sample Proceedings of the VLDB Endowment Paper in LaTeX Format}
\vldbAuthors{Ben Trovato, G. K. M. Tobin, Lars Th{\sf{\o}}rv{$\ddot{\mbox{a}}$}ld, Lawrence P. Leipuner, Sean Fogarty, Charles Palmer, John Smith, Julius P.~Kumquat, and Ahmet Sacan}
\vldbDOI{https://doi.org/10.14778/xxxxxxx.xxxxxxx}
\vldbVolume{12}
\vldbNumber{xxx}
\vldbYear{2019}

\usepackage{textcomp} \usepackage{amsmath,bm}
\usepackage{algorithm}

\usepackage[utf8]{inputenc} 

\usepackage{caption}
\usepackage{microtype}
\usepackage{booktabs}
\usepackage{titlecaps}
\Addlcwords{are or etc in}
\usepackage{xcolor}
\usepackage{xspace}
\usepackage{makecell}
\usepackage{mathtools}
\usepackage{subcaption}
\usepackage{multirow}
\usepackage{algorithmicx}
\usepackage{algpseudocode}
\usepackage{mwe}
\usepackage{multicol}
\usepackage[binary-units,group-separator={,}]{siunitx}
\usepackage{dirtytalk}
\usepackage{url}
\usepackage{pifont}\newcommand{\cmark}{\ding{51}}\newcommand{\xmark}{\ding{55}}\usepackage{tikz}
\usetikzlibrary{shapes}
\newcommand*\circled[1]{\tikz[baseline=(char.base)]{
  \node[shape=circle,draw,inner sep=0.3pt] (char) {#1};}}

\newtheorem{tdef}{Definition}

\DeclarePairedDelimiter\abs             {\lvert}{\rvert}

\definecolor{darkestColor}{HTML}{00BD39} 
\definecolor{darkColor}{HTML}{057D9F}

\begin{document}

\title{Enc\-DB\-DB{}: Searchable Encrypted, Fast, Compressed, In-Memory Database using Enclaves}

\numberofauthors{3} 

\author{
\alignauthor
Benny Fuhry\\
       \affaddr{SAP Security Research}\\
       \affaddr{Karlsruhe, Germany}\\
\alignauthor
Jayanth Jain H A\\
       \affaddr{SAP Security Research}\\
       \affaddr{Karlsruhe, Germany}\\
\alignauthor 
Florian Kerschbaum\\
       \affaddr{University of Waterloo}\\
       \affaddr{Waterloo, Canada}\\
}

\maketitle

\begin{abstract}
Data confidentiality is an important requirement for clients when outsourcing databases to the cloud.
Trusted execution environments, such as Intel SGX, offer an efficient, hardware-based solution to this cryptographic problem.
Existing solutions are not optimized for column-oriented, in-memory databases and pose impractical memory requirements on the enclave.
We present Enc\-DB\-DB{}, a novel approach for client-controlled encryption of a column-oriented, in-memory databases allowing range searches using an enclave.
Enc\-DB\-DB{} offers nine encrypted dictionaries{}, which provide different security, performance and storage efficiency tradeoffs for the data.
It is especially suited for complex, read-oriented, analytic queries, e.g.,\@\xspace{} as present in data warehouses.
The computational overhead compared to plaintext processing is within a millisecond even for databases with millions of entries and the leakage is limited.
Compressed encrypted data requires less space than a corresponding plaintext column.
Furthermore, the resulting code --- and data --- in the enclave is very small reducing the potential for security-relevant implementation errors and side-channel leakages.
\end{abstract}

\section{Introduction}\label{sec:02:Introducton}
Data warehouses are used by companies for business intelligence and decision support.
Such warehouses contain large datasets and the underlying database management systems (DBMS) are optimized for complex, read-oriented, analytic queries.
Outsourcing the data and query processing to the cloud, more specifically to a Database-as-a-Service (DBaaS) provider, can reduce costs, minimize maintenance efforts and increase availability.
However, companies are reluctant to outsource their sensitive data to an untrusted DBaaS provider due to possible
data leakage, government intrusion, and legal liability.

Cryptographic solutions can be a building block for an encrypted cloud database.
For instance, fully homomorphic encryption (FHE)~\cite{gentry_fully_2009} supports arbitrary computations on encrypted data, but is still too slow for practical deployability~\cite{coron_fully_2011,gentry_homomorphic_2012}.
CryptDB~\cite{popa_cryptdb:_2011} and Monomi~\cite{Monomi} use multiple encryption schemes, e.g.,\@\xspace{} probabilistic encryption, deterministic encryption, and order-preserving encryption~\cite{agrawal_order_2004,boldyreva_ope_2009,boldyreva_ope_2011,kerschbaum_optimal_2014} to perform different database functionalities.
The encryption schemes are layered and/or stored in parallel, introducing a storage overhead, and careful query rewriting is necessary to receive a result securely and efficiently.

An alternative approach is to build an encrypted database based on a trusted execution environment (TEE).
TEEs provide an isolated, trusted environment for application code and data known as an \emph{enclave}.
Intel SGX~\cite{Intel_SGX3,costanintel,Intel_SGX2,SGX_Ref,iscaTut,Intel_SGX1}, a TEE that is integrated into (most) modern Intel CPUs, sparked a new wave of research in the direction of TEE-based encrypted databases~\cite{baumann_shielding_2014,fuhry2017hardidx,gribov2017stealthdb,EnclaveDB,ObliDB18}.
SGX enclaves provide isolation against any other code, e.g.,\@\xspace{} application code, other enclaves, and the OS\@.
However, current TEE approaches assume an unrealistic size of enclaves~\cite{baumann_shielding_2014,EnclaveDB}, do not provide DBMS functionality~\cite{fuhry2017hardidx}, do not support persistency~\cite{ObliDB18}, or leak the result of every primitive operation~\cite{gribov2017stealthdb}.
Also, these solutions do not consider data compression to reduce the size of large databases.

We propose and implement Enc\-DB\-DB{}, a high-perfor\-mance, encrypted cloud database supporting analytic queries on large datasets.
We focus on a complex, required query type: range queries.
However, it is straightforward to also support e.g.,\@\xspace{} count, aggregation, and average calculations.
Enc\-DB\-DB{} is based on a column-oriented, dictionary encoding{} based, in-memory database.
Column-oriented data storage optimizes the processing of analytic workloads~\cite{DBS-024,boncz1999mil,copeland1985decomposition,stonebraker2005c}, in-memory processing boosts the overall performance~\cite{dewitt1984implementation,garcia1992main,larson2016modern}, and dictionary encoding{} reduces the storage space overhead of large (encrypted) datasets~\cite{abadi2006integrating,Willhalm}.
Table~\ref{tab:02:teeDatabases} compares our approach to the most relevant related work.

\begin{table*}[t]
    \fontsize{9}{10.5}\selectfont \caption{Comparison of existing TEE based encrypted databases and Enc\-DB\-DB{}. The overheads compare the respective approach with a plaintext database. We present lower bounds of the overheads to the advantage of the approaches, taken from the corresponding papers where available. More details are given in Section~\ref{sec:02:relatedWork}.}
    \centering
    \label{tab:02:teeDatabases}\begin{tabular}{lllllll}
        \toprule
        \multirow{2}[3]{*}{Approach} & \multirow{2}[3]{*}{Workload} & \multirow{2}[3]{*}{Protection Object} & \multirow{2}[3]{*}{\makecell[l]{Comp-\\ression}} & \multicolumn{2}{c}{Overhead} & \multirow{2}[3]{*}{LOC} \\
        \cmidrule(lr){5-6}
        & & & & Storage & Performance & \\
        \midrule
        EnclaveDB~\cite{EnclaveDB}              & OLTP          & in-memory storage and query engine    &  \xmark & N/A                     & \(>\SI{20}{\percent}\)        & \(\sim\)\num{235000} \\
        ObliDB~\cite{ObliDB18}                  & OLTP \& OLAP  & data structure (array or $B^+$-tree\xspace{})   &  \xmark & \(>\SI{100}{\percent}\) & \(>\SI{200}{\percent}\)        & \(\sim\)\num{10000} \\
        StealthDB~\cite{gribov2017stealthdb}    & OLTP          & primitive operators (e.g.,\@\xspace{} \(\leq\), \(\geq\), \(+\), \(*\)) &  \xmark & \(>\SI{300}{\percent}\) & \(>\SI{20}{\percent}\)        & \(\sim\)\num{1500} \\
Enc\-DB\-DB{}                                & OLAP          & data structure (dictionaries{})  &  \cmark & \(<\SI{100}{\percent}\) & \(\sim \SI{8.9}{\percent}\)   & \num{1129} \\
        \bottomrule
    \end{tabular}\end{table*}

The main contributions of Enc\-DB\-DB{} are:
\begin{itemize}
    \setlength{\itemsep}{0pt}
    \item New architecture for search over encrypted data suitable for column-oriented, in-memory data\-bases. 
    \item Nine different encrypted dictionaries{} from which the data owner can freely select on column granularity. They provide different security (order and frequency leakage), performance and storage efficiency tradeoffs. The security ranges from the equivalent of deterministic order-revealing encryption~\cite{boneh2015semantically} to range predicate encryption~\cite{lu_privacy-preserving_2012}.
    \item Integration into MonetDB~\cite{boncz_breaking_2008,boncz2005monetdb,idreos_monetdb:_2012}, an open source DBMS\@. The enclave has only \num{1129} lines of code, reducing the potential for security-relevant implementation errors and side-channel leakages. Query optimization and auxiliary database functionalities, e.g.,\@\xspace{} storage, transaction, and database recovery management still operate without changes to the original code.
    \item Sub-millisecond overhead for encrypted range queries compared to plaintext range queries, on a real-world customer database containing millions of entries. 
    \item Less storage space required for a compressed, encrypted column with the appropriate encrypted dictionary{} than for a plaintext column with the same data.
\end{itemize}

 \section{Background}\label{sec:02:background}

First, Enc\-DB\-DB{} is a column-oriented, dictionary encoding{} based, in-memory database.
Second, a TEE, more specifically Intel SGX, is used to protect and process data stored in the database.
Third, the database data are encrypted with probabilistic authenticated encryption{}.
We review these three concepts in this section.

\subsection{Column-oriented, Dictionary Encoding{}\\based, In-memory Databases}\label{subsec:02:background:database}

\textbf{In-memory database.}
Many commercial and open source DBMS vendors offer in-memory databases for analytical data processing, e.g.,\@\xspace{} SAP HANA~\cite{saphana}, Oracle RDBMS~\cite{oracledatabase}, and MonetDB~\cite{monetdbWebsite}.
In-memory databases permanently store the primary data in main memory and use the disk as secondary storage.
The major benefit of in-memory databases is the lower access time of main memory compared to disk storage.
This speeds up every data access for which disk access would be necessary.
Additionally, it leads to shorter locking times in concurrency control, thus fewer cache flushes and a better CPU utilization.
See~\cite{dewitt1984implementation, garcia1992main, larson2016modern} for more details.

\textbf{Column-oriented, In-memory Database.}
One possible database storage concept is to store the data column-oriented, i.e.,\@\xspace{} successive values of each column are stored consecutively, and surrogate identifiers are (implicitly) introduced to connect the rows~\cite{DBS-024, boncz1999mil, copeland1985decomposition, stonebraker2005c}. 
The combination of in-memory databases and column-oriented storage reduces the number of cache misses, which strongly influences the in-memory performance.
All in-memory databases mentioned above support column-orien\-ted storage.

The main drawbacks of column-oriented storage are: (1) so-called tuple-reconstruction is necessary to re-assemble a projection involving multiple attributes and (2) inserts and updates of a tuple are written to non-contiguous storage locations. 
These problems are not severe in the context of analytical applications, e.g.,\@\xspace{} data warehousing and business intelligence, because analytical queries often involve a scan on a significant amount of all tuples, but only a small subset of all columns~\cite{boncz1999mil,krueger2010enterprise}. 
Additionally, bulk loading of data is often used in this context and complex, long, read-only queries are executed afterwards~\cite{harizopoulos2006performance,stonebraker2005c}.
An example query is a report on total sales per country for products in a certain price range.
Only the few columns that are involved in the query have to be loaded and they can be processed sequentially, which is beneficial as it decreases cache misses of CPUs.

\textbf{Column-oriented, Dictionary Encoding{} based, In-\linebreak[4]memory Databases.}
The three commercial DBMSes mentioned above and many other databases use data compression mechanisms to exploit redundancy within data~\cite{abadi2006integrating,Willhalm}. 
Abadi et al.~\cite{abadi2006integrating} study multiple database compression schemes, e.g.,\@\xspace{} null suppression, run-length encoding and dictionary encoding{}, and show how they can be applied to column-oriented databases.
According to the authors, column-orien\-ted databases particularly profit from compression.
In this paper, we only consider dictionary encoding{}, because it is the most prevalent compression used in column-oriented data\-bases~\cite{abadi2006integrating}.

Throughout this paper, we say a tuple \ensuremath{T{}}\xspace{} contains \ensuremath{\abs{T{}}}\xspace{} values, i.e.,\@\xspace{} \(\ensuremath{T{}}\xspace{} = (\allowbreak{}\ensuremath{T{}_{0}}\xspace,\allowbreak{}\ldots{},\allowbreak{}\ensuremath{T{}_{\ensuremath{\abs{T{}}}\xspace{} - 1}}\xspace )\).
We use \(\ensuremath{v}\xspace{} \in \ensuremath{T{}}\xspace{}\) as a shorthand for a value \ensuremath{v}\xspace{} that is contained in the tuple \ensuremath{T{}}\xspace{}.
The idea of dictionary encoding{} is to split a column \(\ensuremath{C{}}\xspace{} = (\allowbreak{}\ensuremath{C{}_{0}}\xspace,\allowbreak{}\ldots{},\allowbreak{}\ensuremath{C{}_{\ensuremath{\abs{C{}}}\xspace{} - 1}}\xspace )\) into two structures: a \emph{dictionary{}} \ensuremath{D{}}\xspace{} and an \emph{attribute vector{}} \ensuremath{AV{}}\xspace{}.
The dictionary{} \(\ensuremath{D{}}\xspace{} = ( \ensuremath{D{}_{0}}\xspace, \allowbreak \ldots{}, \allowbreak \ensuremath{D{}_{\ensuremath{\abs{D{}}}\xspace - 1}}\xspace )\) is filled with all values \(\ensuremath{v}\xspace{} \in \ensuremath{C{}}\xspace{}\) and every \ensuremath{v}\xspace{} has to be present in \ensuremath{D{}}\xspace{} at least once. 
The index \(i\) of a dictionary{} entry \ensuremath{D{}_{i}}\xspace is called the ValueID{} (\ensuremath{vid{}}\xspace{}) that corresponds to this value.
The attribute vector{} \(\ensuremath{AV{}}\xspace{} = (\allowbreak{} \ensuremath{AV{}_{0}}\xspace, \allowbreak \ldots{}, \allowbreak{} \ensuremath{AV{}_{\ensuremath{\abs{AV{}}}\xspace{} - 1}}\xspace )\) is constructed by replacing all values \(\ensuremath{v}\xspace{} \in \ensuremath{C{}}\xspace{}\) with one \ensuremath{vid{}}\xspace{} that corresponds to \ensuremath{v}\xspace{}.
As a result, \ensuremath{AV{}}\xspace{} contains \(\ensuremath{\abs{AV{}}}\xspace{} = \ensuremath{\abs{C{}}}\xspace{}\) ValueID{}s{}.
The index \(j\) of an entry \ensuremath{AV{}_{j}}\xspace is called its RecordID{} (\ensuremath{rid{}}\xspace{}).
\ensuremath{un(C{})}\xspace{} denotes the set of unique values in \ensuremath{C{}}\xspace{}, \ensuremath{\abs{un(C{})}}\xspace{} the amount of unique values, \ensuremath{oc(C{},v{}}\xspace){} the occurrence indices of a unique value \ensuremath{v}\xspace{} in \ensuremath{C{}}\xspace{}, and \ensuremath{\abs{oc(C{},v{})}}\xspace{} the number of occurrences of \ensuremath{v}\xspace{}.
We define the correctness of a column split as follows:
\begin{tdef}[Split Correctness]\label{def:2:split_cor} Given a column \ensuremath{C{}}\xspace{}, we say that a split of \ensuremath{C{}}\xspace{} into a dictionary{} \ensuremath{D{}}\xspace{} and an attribute vector{} \ensuremath{AV{}}\xspace{} is correct if \(i\) is the ValueID{} stored in the attribute vector{} at position \(j\) and \ensuremath{D{}_{i}}\xspace equals \ensuremath{C{}_{j}}\xspace, i.e.,\@\xspace{} \(\forall{} j \in [0, \ensuremath{\abs{AV{}}}\xspace{} - 1] \colon{} \allowbreak i = \ensuremath{AV{}_{j}}\xspace \wedge{} \ensuremath{D{}_{i}}\xspace = \ensuremath{C{}_{j}}\xspace\).
\end{tdef}

\begin{figure}[h]
  \centering
  \includegraphics[width=\linewidth]{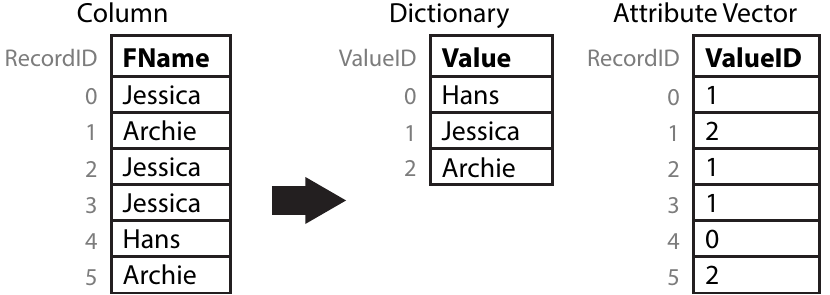}
  \caption{Dictionary encoding{} example}\label{fig:02:dictionaryEx}
\end{figure}

In Figure~\ref{fig:02:dictionaryEx}, we present a split example based on a small first name column (\texttt{FName}).
For instance, Jessica was inserted in the dictionary{} at the ValueID{} \(1\) and all positions from the original column that contained Jessica are replaced by this ValueID{} in the attribute vector{} (see RecordID{}s{} \(0\), \(2\) and \(3\)).
The set of unique values is \(\ensuremath{un(C{})}\xspace{} = \{\text{Hans}, \text{Jessica},\allowbreak{}\text{Archie}\}\) and Archie occurs at the positions \(\ensuremath{oc(C{},\text{Archie}}\xspace) = \{1, 5\}\).

Note that a split column requires less space than the original column in many cases, because a ValueID{} of \(i\) Bits is sufficient to represent \(2^i\) different values in the attribute vector{} and the (variable-length) values only have to be stored once in the dictionary{}.
For instance, a column that contains \num{10000} strings of 10 characters each, but only \(256\) unique values, requires \(256 \cdot \SI{10}{\byte}\) for the dictionary and \(\num{10000} \cdot \SI{1}{\byte}\) for the attribute vector{}.
In total, dictionary encoding{} reduces the required storage from \SI{100000}{\byte} to \SI{12650}{\byte}.
Dictionary encoding{} has the best compression rate if columns contain few unique but many frequent values, because every value has to be stored only once.
The real-world data used for our evaluation (see Section~\ref{subsec:02:evaluation:perfEval}) and other studies~\cite{muller2014adaptive,lemke_speeding_2010} show that this is a characteristic of many columns in data warehouses.
High compression rates achieved by dictionary encoding{} sparingly use the scarce resource of in-memory databases --- main memory.

A search for all entries falling in a range \(\ensuremath{R}\xspace{}\) is performed in two steps if dictionary encoding{} is used: a dictionary{} search followed by an attribute vector{} search.
The dictionary{} search \ignorespaces checks for every \(\ensuremath{v}\xspace{} \in \ensuremath{D{}}\xspace{}\) if it falls into \ensuremath{R}\xspace{} and returns the matching ValueID{}s{} (\ensuremath{\mathbf{vid{}}}\xspace{}).
The attribute vector{} search \ignorespaces linearly scans the attribute vector{} searching for every value \(\ensuremath{v}\xspace{} \in \ensuremath{\mathbf{vid{}}}\xspace{}\) and returns a list of matching RecordID{}s{} (\ensuremath{\mathbf{rid{}}}\xspace{}).
This operation is parallelizable with a speedup expected to be linear in the number of threads.

In the example of Figure~\ref{fig:02:dictionaryEx}, a search for \(\ensuremath{R}\xspace{} = [\text{Archie}, \text{Hans}] \) in the dictionary{} returns \(\ensuremath{\mathbf{vid{}}}\xspace{} = \{0,2\}\).
Searching these ValueID{}s{} in the attribute vector{} returns \(\ensuremath{\mathbf{rid{}}}\xspace{} = \{1, 4, 5\}\).

\subsection{Intel Software Guard Extensions (SGX)}\label{subsec:02:background:sgx}

Intel SGX is an instruction set extension that is available in Intel Core processors since the Skylake generation and in Intel Xeon processors since the Kaby Lake generation, making it a widely available TEE\@.
It provides a secure, isolated processing area, called enclave, which guarantees confidentiality and integrity protection to code and data in it, even in an untrusted environment. 
We present SGX's features used by Enc\-DB\-DB{}.
See~\cite{Intel_SGX3,costanintel,Intel_SGX2,SGX_Ref,iscaTut,Intel_SGX1} for more details.

\textbf{Memory Isolation.} 
SGX v2 dedicates \SI{128}{\mega\byte} of the system's main memory (RAM) for the so-called Processor Reserved Memory (PRM).
All code and data in the PRM is encrypted while residing outside of the CPU, and decrypted and integrity checked when the data is loaded into the CPU\@.
All other software on the system, including privileged software such as OS, hypervisor, and firmware, cannot access the PRM\@.
Only about \SI{96}{\mega\byte} of the PRM can be used for enclave code and data, even if multiple enclaves are present.
The OS can swap out enclave pages and SGX ensures integrity, confidentiality and freshness of swapped-out pages, but paging comes with a major performance overhead.

Every program using SGX consists of an enclave and an untrusted part.
The untrusted part is executed as an ordinary process within the virtual memory address space and the enclave memory is mapped into the virtual memory of the untrusted host process.
This mapping allows the enclave to access the entire virtual memory of its host process, while the host process can invoke the enclave only through a well-defined interface.

\textbf{Attestation.} 
SGX has a remote attestation feature, which allows verification of code integrity and authenticity on a remote system.
This is done by hashing (called \emph{measuring} in SGX terminology) the initial code and data loaded into the enclave.
The authenticity of the measurement, as well as the fact that the measurement originates from a benign enclave, is ensured by SGX's attestation feature (refer to~\cite{Intel_SGX3} for details). 
The measurement can be provided to an external party to prove the correct creation of an enclave.
Furthermore, the remote attestation feature allows establishment of a secure channel between an external party and an enclave.
This secure channel can be used to deploy sensitive data, e.g.,\@\xspace{} cryptographic keys, directly into the enclave.

\subsection{Probabilistic Authenticated Encryption{}}\label{subsec:02:background:authEnc}

A probabilistic authenticated encryption{} (\ensuremath{\mathtt{PAE{}}}\xspace{}) scheme provides confidentiality, integrity, and authenticity of encrypted data.
\ensuremath{\mathtt{PAE{}\_}\allowbreak{}\mathtt{Enc}}\xspace{} takes a secret key \ensuremath{SK}\xspace{}, a random initialization vector \ensuremath{IV}\xspace{} and a plaintext value \ensuremath{v}\xspace{} as input and returns a ciphertext \ensuremath{c{}}\xspace{}.
\ensuremath{\mathtt{PAE{}\_}\allowbreak{}\mathtt{Dec}}\xspace{} takes \ensuremath{SK}\xspace{} and \ensuremath{c{}}\xspace{} as input and returns \ensuremath{v}\xspace{} iff \ensuremath{v}\xspace{} was encrypted with \ensuremath{\mathtt{PAE{}\_}\allowbreak{}\mathtt{Enc}}\xspace{} under the initialization vector \ensuremath{IV}\xspace{} and the secret key \ensuremath{SK}\xspace{}.
AES-128 in GCM mode~\cite{dworkin2007recommendation} can be used as a \ensuremath{\mathtt{PAE{}}}\xspace{} implementation. \section{High Level Design of Enc\-DB\-DB{}}\label{sec:02:designOverview}

In this section, we give an overview of Enc\-DB\-DB{}'s setup and query phase, followed by the considered attacker model.

\subsection{Enc\-DB\-DB{} Overview}\label{subsec:02:designOverview:overview}

Enc\-DB\-DB{} provides nine encrypted dictionaries{} and also supports plaintext dictionaries{}.
In the setup phase, one of these is selected per column of the data owner{}'s dataset.
The selection determines how each column is split into a dictionary{} and an attribute vector{}.
All values in encrypted dictionaries{} are encrypted with \ensuremath{\mathtt{PAE{}}}\xspace{} under a key determined by the data owner{}.
The encrypted dictionaries{} provide different tradeoffs regarding security, performance, and storage efficiency.
Enc\-DB\-DB{} is able to process all dictionary{} types together, even if they are mixed in one table.
For brevity, plaintext dictionaries{} are not discussed any further, but performance measurements are shown in the evaluation section.

\begin{figure}[h]
  \centering
  \includegraphics[width=\linewidth]{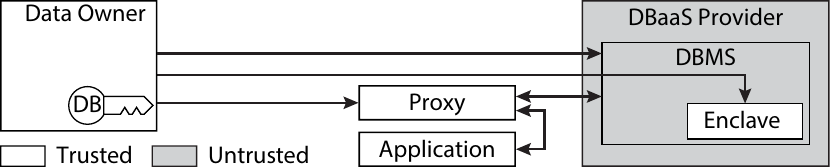}
  \caption{High level design of Enc\-DB\-DB{}}
  \label{fig:02:overview}
\end{figure}

The data owner{}'s dataset is deployed at a DBaaS provider that supports Intel SGX (see Figure~\ref{fig:02:overview}).
SGX can be replaced by any other TEE that provides the required capabilities such as integrity and confidentiality protection of code and data, remote attestation, and secure data provisioning.
Additionally, the data owner{}'s secret key (\ensuremath{SK_{DB}}\xspace{}) is deployed into the SGX enclave that is part of the DBMS and to a trusted proxy{}.

After this setup, the query phase starts, and an application{} can send queries to the DBaaS offering.
These queries are routed through the proxy{}, where all values are encrypted with \ensuremath{\mathtt{PAE{}}}\xspace{} and forwarded to the DBMS, which processes the queries in a pipeline.
The pipeline outputs encrypted range queries on individual columns.
The enclave is used for protected dictionary{} searches.
Attribute vector{} searches and all other DBMS functionality are performed outside of the enclave.
The DBMS combines the individual results into a total result and returns it to the proxy{}.
At the proxy, the total result is decrypted and forwarded to the application{} for which the whole process is transparent.

We introduce the encrypted dictionaries{} in detail in Section~\ref{subsec:02:main:encDTs}.
In Section~\ref{subsec:02:main:designDetail}, we provide an in-depth description of how the encrypted dictionaries{} are used as the main building block for implementing a protected DBMS that supports auxiliary DBMS functions.
In Section~\ref{subsec:02:main:dynData}, we explain how Enc\-DB\-DB{} can handle data insertions, deletions, and updates.

\subsection{Attacker Model}\label{subsec:02:designOverview:attackerModel}

Our attacker model considers the data owner{}, application{} and proxy{} as trusted.
On the server side, we assume an honest-but-curious attacker, i.e.,\@\xspace{} a passive attacker who follows the protocol, but tries to gain as much information as possible.

The enclave code and data is protected by the TEE\@.
The code is assumed not to have intentional data leakage.
However, the attacker can observe all other software running at the DBaaS provider, e.g.,\@\xspace{} the OS, the firmware and the DBMS\@.
As a result, the attacker has full access to data stored on disk and main memory, and she is able to observe the access pattern to them.
Additionally, she can track all communication between the enclave and resources outside of it, and all network communication between the proxy and the DBMS\@.
Note that this includes the incoming queries in which only the data values are encrypted.

Various research studies have shown that SGX is vulnerable to various side-channel attacks, e.g.,\@\xspace{} cache attacks~\cite{brasser2017software}, timing attacks~\cite{weichbrodt2016asyncshock} or page faults~\cite{xu2015controlled}.
Other researchers have presented solutions to these problems.
For instance, how to mitigate the page fault side-channel~\cite{shinde2016preventing}, how to detect side-channels~\cite{chen2017detecting} and how to protect against cache-based side-channels~\cite{gruss2017strong}.
We consider the side-channel exploitation and protection as an orthogonal problem and thus do not consider them in this work.
However, we design our system to have minimal enclave code and therefore, the protections should be straightforward to integrate.
Hardware attacks and Denial of Service (DoS) are out of scope.

We assume that the attacker targets each database column independently, i.e.,\@\xspace{} she does not use correlation information to target columns.
It remains future work to evaluate how decorrelation of columns protects the database in practice.

 \section{Enc\-DB\-DB{} Design}\label{sec:02:main}

In this section, we continue the description of Enc\-DB\-DB{}, which was already introduced in Section~\ref{subsec:02:designOverview:overview}. We first explain the nine encrypted dictionaries{} that Enc\-DB\-DB{} supports and then elaborate how they can be used to build an encrypted DBMS\@.

\subsection{\titlecap{encrypted dictionaries{}}}\label{subsec:02:main:encDTs}

The encrypted dictionaries{} differ from each other in two dimensions --- repetition and order of values in \ensuremath{D{}}\xspace{} --- with three options each (see Table~\ref{tab:02:encDTsCharacteristics}).
The repetition options are: frequency revealing{}, frequency smoothing{}, and frequency hiding{}.
The order options are: sorted lexicographically, sorted and rotated around a random offset, and unsorted.
An encrypted dictionary{} is defined by one option from each dimension, which leads to nine data structures with different security, search time and storage efficiency features.

\begingroup
\setlength{\tabcolsep}{3.5pt}
\begin{table}[htbp]
    \fontsize{9}{10.5}\selectfont \caption{Characteristics of encrypted dictionaries{}}\label{tab:02:encDTsCharacteristics}\centering
    \begin{tabular}{llccc}        
        \toprule        
         & & \multicolumn{3}{c}{order options}\\
        \cmidrule(lr){3-5}
         & & sorted          & rotated & unsorted\\
        \midrule
\parbox[b]{7mm}{\multirow{3}{*}{\rotatebox[origin=c]{90}{\makecell{repetition\\options}}}} &
        frequency revealing{}     & ED1 & ED2 & ED3\\[4pt]
        & frequency smoothing{}   & ED4 & ED5 & ED6\\[4pt]
        & frequency hiding{}   & ED7 & ED8 & ED9\\        
        \bottomrule
    \end{tabular}\end{table}
\endgroup

The idea of the repetition options is to increase the number of repetitions of dictionary{} values from frequency revealing{} to frequency hiding{}.
This directly influences two features of the resulting encrypted dictionaries{} (see Table~\ref{tab:02:encDTsRepetition}): the security feature frequency leakage and dictionary{} size (\ensuremath{\abs{D{}}}\xspace{}).
Note that \ensuremath{\abs{D{}}}\xspace{} is fixed for frequency revealing{} and frequency hiding{}.
For frequency smoothing{}, the worst-case size is \ensuremath{\abs{AV{}}}\xspace{}, but we give the average size, which depends on a configurable parameter \ensuremath{bs{}_{\max}}\xspace{}.

\begingroup
\setlength{\tabcolsep}{4pt}
\begin{table}[htbp]
    \fontsize{9}{10.5}\selectfont \caption{Security feature frequency leakage and dictionary size of repetition options}\label{tab:02:encDTsRepetition}\centering
    \begin{tabular}{lll}
        \toprule
        repetition options  & frequency leak. & dictionary size \ensuremath{\abs{D{}}}\xspace{}\\
        \midrule
        frequency revealing{}     & full              & \ensuremath{\abs{un(C{})}}\xspace{}\\
        frequency smoothing{}     & bounded           & \(\sim \sum_{\ensuremath{v}\xspace{} \in \ensuremath{C{}}\xspace{}}{\frac{2 \cdot \ensuremath{\abs{oc(C{},v{})}}\xspace{}}{1 + \ensuremath{bs{}_{\max}}\xspace{}}}\)\\
        frequency hiding{}     & none              & \ensuremath{\abs{AV{}}}\xspace{}\\
        \bottomrule
    \end{tabular}\end{table}
\endgroup

\begin{figure*}[t]
    \setlength{\belowcaptionskip}{-4pt}
    \begin{subfigure}[c]{0.14\textwidth}    
        \includegraphics[]{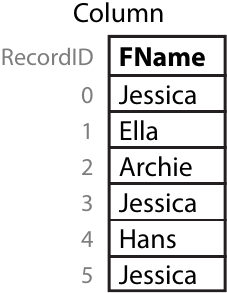}
        \subcaption{Column \ensuremath{C{}}\xspace{}}    
    \end{subfigure}\hfill \begin{subfigure}[c]{0.28\textwidth}    
        \includegraphics[]{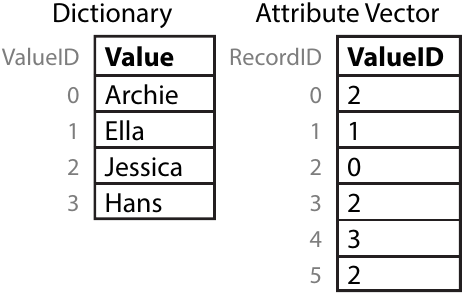}
        \subcaption{ED1}    
    \end{subfigure}\hfill \begin{subfigure}[c]{0.28\textwidth}    
        \includegraphics[]{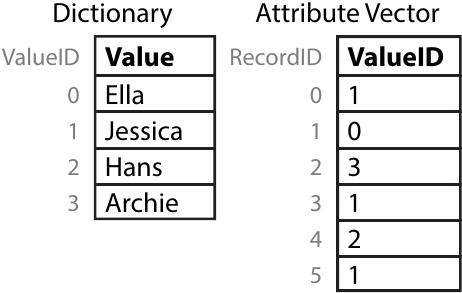}
        \subcaption{ED2 with \(\textit{rndOffset}{} = 3\)}    
    \end{subfigure}\hfill \begin{subfigure}[c]{0.28\textwidth}    
        \includegraphics[]{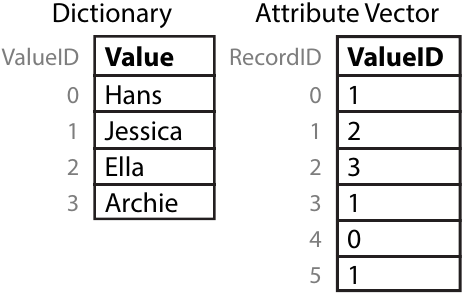}
        \subcaption{ED3}    
    \end{subfigure}
    \caption{(a) example column \ensuremath{C{}}\xspace{} processed with (b) ED1, (c) ED2, and (d) ED3 before encryption}\label{fig:02:exDT1toDT3}
\end{figure*}

The order options also determine two features of the encrypted dictionaries{} (see Table~\ref{tab:02:encDTsOrder}).
First, they determine the security feature order leakage, i.e.,\@\xspace{} the information an attacker with memory access can learn about the plaintext order of the encrypted values in \ensuremath{D{}}\xspace{}.
Second, they determine the search time combining the dictionary{} and attribute vector{} search time.
The dictionary{} search time depends on \ensuremath{\abs{D{}}}\xspace{} and the search algorithm, which differs for the order options.
The attribute vector{} search time depends on the amount of ValueID{}s{} returned by the dictionary{} search, because \ensuremath{AV{}}\xspace{} has to be scanned for them.

\begin{table}[htbp]
    \fontsize{9}{10.5}\selectfont \caption{Security feature order leakage and search time of order options}\label{tab:02:encDTsOrder}\centering
    \begin{tabular}{lll}
        \toprule
        order options   & order leakage & search time\\
        \midrule
        sorted          & full          & \(O(\log \ensuremath{\abs{D{}}}\xspace{}) + O(\ensuremath{\abs{AV{}}}\xspace{})\)\\
        rotated         & bounded       & \(O(\log \ensuremath{\abs{D{}}}\xspace{}) + O(\ensuremath{\abs{AV{}}}\xspace{})\)\\
        unsorted        & none          & \(O(\ensuremath{\abs{D{}}}\xspace{}) + O(\ensuremath{\abs{AV{}}}\xspace{}\cdot\ensuremath{\mathbf{\abs{vid{}}}}\xspace{})\)\\
        \bottomrule
    \end{tabular}\end{table}

Three operations differ for the nine encrypted dictionaries{}: (1) creation of the encrypted dictionaries{}, (2)  dictionary{} search inside the enclave at the DBaaS provider, and (3) attribute vector{} search in the untrusted realm at the DBaaS provider.
In the next sections, we denote the corresponding operations as (1) \ensuremath{\mathtt{EncDB}}\xspace{}, (2) \ensuremath{\mathtt{EnclDict}\-\mathtt{Search}}\xspace{}, and (3) \ensuremath{\mathtt{AttrVect}\-\mathtt{Search}}\xspace{}, and describe these operations in detail.

As mentioned before, an encrypted dictionary{} is defined by an order and a repetition option.
We start by describing the frequency revealing{} algorithm and then explain how it is combined with the three order options to instantiate ED1--ED3.
Then, we do the same for the frequency smoothing{} algorithm and its combinations (ED4--ED6) followed by the frequency hiding{} algorithm and its combinations (ED7--ED9).
We always assume a closed search range.
Open or half-open ranges can be handled trivially, however we omit the details to provide a concise description.

\subsubsection*{Frequency revealing{}}

In the frequency revealing{} algorithm, the split of a column \ensuremath{C{}}\xspace{} is performed by inserting each unique value \(\ensuremath{v}\xspace{} \in \ensuremath{un(C{})}\xspace{}\) into \ensuremath{D{}}\xspace{} exactly once at an arbitrary position, i.e.,\@\xspace{} \(\ensuremath{\abs{D{}}}\xspace{} = \ensuremath{\abs{un(C{})}}\xspace{} \wedge \forall \ensuremath{v}\xspace{} \in \ensuremath{un(C{})}\xspace{} \colon \ensuremath{v}\xspace{} \in \ensuremath{D{}}\xspace{}\).
The ValueID{}s{} in \ensuremath{AV{}}\xspace{} are set such that the split is correct according to Definition~\ref{def:2:split_cor}.

The frequency revealing{} algorithm provides the best compression rate that is possible with dictionary encoding{} and thus is the most storage efficient repetition option.
However, an attacker can learn the frequency of each value \(\ensuremath{D{}_{j}}\xspace \in \ensuremath{D{}}\xspace{}\) by counting the occurrences of \(j\) in \ensuremath{AV{}}\xspace{}.
This is still true if each \(\ensuremath{v}\xspace{} \in{} \ensuremath{D{}}\xspace{}\) is encrypted with probabilistic authenticated encryption{}.
Therefore, the three frequency revealing{} encrypted dictionaries{} presented next have full frequency leakage.

\textbf{ED1.} 
For each column \ensuremath{C{}}\xspace{} that is protected with ED1, \ensuremath{\mathtt{EncDB\_}\allowbreak{}\mathtt{1}}\xspace performs the split operation according to the frequency revealing{} algorithm, sorts the values \(\ensuremath{v}\xspace{} \in \ensuremath{D{}}\xspace{}\) lexicographically
and adjusted the ValueID{}s{} in \ensuremath{AV{}}\xspace{} such that the split is correct.
Afterwards, \ensuremath{\mathtt{EncDB\_}\allowbreak{}\mathtt{1}}\xspace derives \ensuremath{SK_{D}}\xspace{} from the data owner{}'s secret key \ensuremath{SK_{DB}}\xspace{}, the table name, and the column name.
It then encrypts all values \ensuremath{v}\xspace{} individually with \ensuremath{\mathtt{PAE{}\_}\allowbreak{}\mathtt{Enc}}\xspace{} under \ensuremath{SK_{D}}\xspace{} and a random initialization vector \ensuremath{IV}\xspace{}.
The resulting dictionary{} containing encrypted values is denoted as \ensuremath{eD{}}\xspace{}.
Figure~\ref{fig:02:exDT1toDT3} (a) presents an example column \ensuremath{C{}}\xspace{} and Figure~\ref{fig:02:exDT1toDT3} (b) the result of ED1 before \ensuremath{\mathtt{PAE{}\_}\allowbreak{}\mathtt{Enc}}\xspace{} is performed.
ED1 has full order leakage because an attacker knows the plaintext order of the encrypted values \(\ensuremath{c{}}\xspace{} \in{} \ensuremath{eD{}}\xspace{}\).

ED1's dictionary{} search (\ensuremath{\mathtt{EnclDict}\-\mathtt{Search\_}\allowbreak{}\mathtt{1}}\xspace), which is executed in the enclave at the DBaaS provider, is presented in Algorithm~\ref{algo:02:dictSearch_1}.
The function gets an encrypted range \ensuremath{\tau}\xspace{} and an encrypted dictionary{} \ensuremath{eD{}}\xspace{} as input.
First it derives \ensuremath{SK_{D}}\xspace{} and decrypts the start and end of the range individually.
Then, one leftmost and one rightmost binary search is performed to find the dictionary{} indices where the searched range starts (\ensuremath{vid{}_{min}}\xspace) and ends (\ensuremath{vid{}_{max}}\xspace).
All dictionary{} values are encrypted and stored in untrusted memory.
Thus, the binary searches load the values into the enclave individually, decrypt them there, and compare them with the search value.
The number of load, decrypt and compare operations is logarithmic in \ensuremath{\abs{D{}}}\xspace{}.
In our implementation we use the result of the searches, and whether or not a value was found, to handle cases in which a value is not present.
This detail is omitted for brevity in Algorithm~\ref{algo:02:dictSearch_1}.

\begin{algorithm}		
    \fontsize{8}{9.5}\selectfont \begin{algorithmic}[1]
        \State \(\ensuremath{SK_{D}}\xspace{} = \Call{DeriveKey}{\ensuremath{SK_{DB}}\xspace{},colName,tabName}\)
        \State \ensuremath{R}\xspace{} = (\ensuremath{R_s}\xspace{}, \ensuremath{R_e}\xspace{}) = \ensuremath{\big(\ensuremath{\mathtt{PAE{}\_}\allowbreak{}\mathtt{Dec}(\allowbreak{}\ensuremath{SK_{D}}\xspace{},\allowbreak{}\ensuremath{\tau_s}\xspace{})}\xspace,\ensuremath{\mathtt{PAE{}\_}\allowbreak{}\mathtt{Dec}(\allowbreak{}\ensuremath{SK_{D}}\xspace{},\allowbreak{}\ensuremath{\tau_e}\xspace{})}\xspace\big)}
        \State \ensuremath{vid{}_{min}}\xspace = \Call{BinarySearchLM}{\ensuremath{eD{}}\xspace{},\ensuremath{R_s}\xspace{}}
        \State \ensuremath{vid{}_{max}}\xspace = \Call{BinarySearchRM}{\ensuremath{eD{}}\xspace{},\ensuremath{R_e}\xspace{}}

        \State \textbf{return} \ensuremath{\mathbf{vid{}}}\xspace{} = (\ensuremath{vid{}_{min}}\xspace, \ensuremath{vid{}_{max}}\xspace)
    \end{algorithmic}
    \caption{\ensuremath{\mathtt{Encl\allowbreak{}Dict\allowbreak{}Search\_}\-\mathtt{1}(\allowbreak{}\ensuremath{\tau}\xspace{},\ensuremath{eD{}}\xspace{})}\xspace\label{algo:02:dictSearch_1}}
\end{algorithm}

Note that only very small, constant enclave memory is required for \ensuremath{\mathtt{EnclDict}\-\mathtt{Search\_}\allowbreak{}\mathtt{1}}\xspace as well as for the \ensuremath{\mathtt{EnclDict}\-\mathtt{Search}}\xspace{} operations of all other encrypted dictionaries{}.
Especially, the required enclave memory is independent of \ensuremath{\abs{D{}}}\xspace{}.

Afterwards, \ensuremath{\mathtt{AttrVect}\-\mathtt{Search\_}\allowbreak{}\mathtt{1}}\xspace is executed in the untrusted realm at the DBaaS provider. 
It linearly scans the corresponding \ensuremath{AV{}}\xspace{}, checks if the ValueID{}s{} fall between \ensuremath{vid{}_{min}}\xspace and \ensuremath{vid{}_{max}}\xspace, and returns the matching RecordID{}s{} (\ensuremath{\mathbf{rid{}}}\xspace{}), i.e.,\@\xspace{} \(\ensuremath{\mathbf{rid{}}}\xspace{} = \{ i \, | \, \ensuremath{AV{}_{i}}\xspace \in{} \ensuremath{AV{}}\xspace{} \wedge \ensuremath{AV{}_{i}}\xspace \in [\ensuremath{vid{}_{min}}\xspace, \ensuremath{vid{}_{max}}\xspace]\} \).
This operation is parallelizable with a speedup expected to be linear in the number of threads.

\textbf{ED2.} 
The idea of the rotated algorithm, which is used in ED2, is to sort and randomly rotate \ensuremath{D{}}\xspace{}.
\ensuremath{\mathtt{EncDB\_}\allowbreak{}\mathtt{2}}\xspace executes the frequency revealing{} algorithm, sorts the values in \ensuremath{D{}}\xspace{} lexicographically, performs a random rotation of \ensuremath{D{}}\xspace{} as explained in the following paragraph, adjusts the ValueID{}s{} in \ensuremath{AV{}}\xspace{} such that the split is correct, and encrypts all \(\ensuremath{v}\xspace{} \in \ensuremath{D{}}\xspace{}\) with \ensuremath{\mathtt{PAE{}}}\xspace{} under \ensuremath{SK_{D}}\xspace{} and a random \ensuremath{IV}\xspace{}, resulting in \ensuremath{eD{}}\xspace{}.

\ensuremath{\mathtt{EncDB\_}\allowbreak{}\mathtt{2}}\xspace generates a random offset (\emph{\textit{rndOffset}{}}) and rotates \ensuremath{D{}}\xspace{} by this value.
More formally, let \ensuremath{D'{}}\xspace{} be the sorted dictionary{}, then \(\ensuremath{D{}}\xspace{} = (\ensuremath{D{}_{i}}\xspace \, | \, \ensuremath{D{}_{i}}\xspace = \ensuremath{D'{}_{j}}\xspace \wedge{} i = (j + \text{\textit{rndOffset}{}}) \bmod{} \ensuremath{\abs{D'{}}}\xspace)\).
\ensuremath{\mathtt{EncDB\_}\allowbreak{}\mathtt{2}}\xspace encrypts \textit{rndOffset}{} with \ensuremath{\mathtt{PAE{}}}\xspace{} under \ensuremath{SK_{D}}\xspace{} and a random \ensuremath{IV}\xspace{}, and attaches the resulting \emph{\textit{encRndOffset}{}} to \ensuremath{eD{}}\xspace{}.

Figure~\ref{fig:02:exDT1toDT3} (c) illustrates an example with \(\textit{rndOffset}{} = 3\) (before encryption).
For instance, \say{Jessica} has the ValueID{} \(2\) in a sorted dictionary{} \ensuremath{D'{}}\xspace{}.
After the rotation, the ValueID{} is \(1 =(2 + 3) \bmod{} 4\).

The order leakage is bounded, because an attacker who can observe no or a limited number of queries, does not know where the smallest and largest values are stored in \ensuremath{eD{}}\xspace{}.
The idea of modular order-preserving encryption in the context of probabilistic encryption was introduced in~\cite{Kerschbaum17}.

The processing inside the enclave (\ensuremath{\mathtt{EnclDict}\-\mathtt{Search\_}\allowbreak{}\mathtt{2}}\xspace) is illustrated in Algorithm~\ref{algo:02:dictSearch_2}.
First, \ensuremath{SK_{D}}\xspace{} is derived, and the encrypted range \ensuremath{\tau}\xspace{} and \textit{encRndOffset}{} are decrypted with it.
Then, a special variant of binary search, which is explained in the next paragraph, is called to search the start and the end of the range --- \ensuremath{vid{}_{min}}\xspace and \ensuremath{vid{}_{max}}\xspace.
These indices have to be processed further inside of the enclave, because the positions of the indices relative to \textit{rndOffset}{} define the final result of the dictionary{} search and \textit{rndOffset}{} is sensitive.
There are three possibilities: both indices are lower than \textit{rndOffset}{}; both are greater than or equal to \textit{rndOffset}{}; or \ensuremath{vid{}_{min}}\xspace is above and \ensuremath{vid{}_{max}}\xspace is below \textit{rndOffset}{}.
In the first and second case, the results are in the range \ensuremath{\big(\ensuremath{vid{}_{min}}\xspace, \ensuremath{vid{}_{max}}\xspace\big)}.
In the third case, there are again two possibilities: \ensuremath{vid{}_{min}}\xspace does or does not equal \ensuremath{\abs{eD{}}}\xspace{}.
In the first case, the range start was not found in \ensuremath{eD{}}\xspace{}, but it is higher than the last value in it.
Accordingly, all results are in the range \ensuremath{\big(0, \ensuremath{vid{}_{max}}\xspace\big)}.
Otherwise, the results are split in a lower range \ensuremath{\big(0, \ensuremath{vid{}_{max}}\xspace\big)} and an upper range \ensuremath{\big(\ensuremath{vid{}_{min}}\xspace, \ensuremath{\abs{eD{}}}\xspace{} - 1\big)}.
We always return a dummy range if the result is only one range to simplify attribute vector{} search.

\begin{algorithm}		
    \fontsize{8}{9.5}\selectfont \begin{algorithmic}[1]
        \State \(\ensuremath{SK_{D}}\xspace{} = \Call{DeriveKey}{\ensuremath{SK_{DB}}\xspace{},colName,tabName}\)
        \State \(\ensuremath{R}\xspace{} = (\ensuremath{R_s}\xspace{}, \ensuremath{R_e}\xspace{}) = \ensuremath{\big(\ensuremath{\mathtt{PAE{}\_}\allowbreak{}\mathtt{Dec}(\allowbreak{}\ensuremath{SK_{D}}\xspace{},\allowbreak{}\ensuremath{\tau_s}\xspace{})}\xspace,\ensuremath{\mathtt{PAE{}\_}\allowbreak{}\mathtt{Dec}(\allowbreak{}\ensuremath{SK_{D}}\xspace{},\allowbreak{}\ensuremath{\tau_e}\xspace{})}\xspace\big)}\)
        \State \(\textit{rndOffset}{} = \ensuremath{\mathtt{PAE{}\_}\allowbreak{}\mathtt{Dec}(\allowbreak{}\ensuremath{SK_{D}}\xspace{},\allowbreak{}\textit{encRndOffset}{})}\xspace\)
        \State \ensuremath{vid{}_{min}}\xspace = \Call{BinSearchSpecialS}{\ensuremath{eD{}}\xspace{},\ensuremath{R_s}\xspace{},\textit{rndOffset}{},\ensuremath{SK_{D}}\xspace{}}
        \State \ensuremath{vid{}_{max}}\xspace = \Call{BinSearchSpecialE}{\ensuremath{eD{}}\xspace{},\ensuremath{R_e}\xspace{},\textit{rndOffset}{},\ensuremath{SK_{D}}\xspace{}}
        \State \ensuremath{\mathbf{vid{}}}\xspace{} = \(\emptyset{}\)
        \If{(\ensuremath{vid{}_{min}}\xspace \(<\) \textit{rndOffset}{} \(\&\) \ensuremath{vid{}_{max}}\xspace \(<\) \textit{rndOffset}{}) \(|\) \\ \ \ \ (\ensuremath{vid{}_{min}}\xspace \(\geqslant\) \textit{rndOffset}{} \(\&\) \ensuremath{vid{}_{max}}\xspace \(\geqslant\) \textit{rndOffset}{})}
            \State \ensuremath{\mathbf{vid{}}}\xspace{} = \{\ensuremath{\big(\ensuremath{vid{}_{min}}\xspace, \ensuremath{vid{}_{max}}\xspace\big)}, \ensuremath{\big(-1, -1\big)}\}
        \ElsIf{\ensuremath{vid{}_{min}}\xspace \(\geqslant\) \textit{rndOffset}{} \(\&\) \ensuremath{vid{}_{max}}\xspace \(<\) \textit{rndOffset}{}}         
            \If{\ensuremath{vid{}_{min}}\xspace \(!=\) \ensuremath{\abs{eD{}}}\xspace{}}
            \State \ensuremath{\mathbf{vid{}}}\xspace{} = \{\ensuremath{\big(0, \ensuremath{vid{}_{max}}\xspace\big)}, \ensuremath{\big(\ensuremath{vid{}_{min}}\xspace, \ensuremath{\abs{eD{}}}\xspace{} - 1\big)}\}
            \Else
            \State \ensuremath{\mathbf{vid{}}}\xspace{} = \{\ensuremath{\big(0, \ensuremath{vid{}_{max}}\xspace\big)}, \ensuremath{\big(-1, -1\big)}  \}
            \EndIf            
        \EndIf

        \State \textbf{return} \ensuremath{\mathbf{vid{}}}\xspace{}
    \end{algorithmic}
    \caption{\ensuremath{\mathtt{Encl\allowbreak{}Dict\allowbreak{}Search\_}\-\mathtt{2}(\allowbreak{}\ensuremath{\tau}\xspace{},\ensuremath{eD{}}\xspace{},\textit{encRndOffset}{})}\xspace\label{algo:02:dictSearch_2}}
\end{algorithm}

Algorithm~\ref{algo:02:BinarySearchSpecial} presents the details of the special binary search with slightly different handling of the range start and end.
The goal is to perform a binary search that has an access pattern that is independent of \textit{rndOffset}{}.
A binary search that simply considers \textit{rndOffset}{} during the data access would leak \textit{rndOffset}{} in the first round, which would completely thwart the additional protection.

The algorithm uses a string encoding operation (\texttt{ENCODE}), which converts string values of a fixed maximal length to an integer representation preserving the lexicographical data order. 
Each character is converted individually to an integer of fixed length and the integers are concatenated to one resulting integer.
For instance, the encoding of \say{AB} would be \(3334\) and \say{BA} would lead to \(3433\).
The lexicographical order is preserved by right padding the resulting integer to a fixed maximal length.
In many DBMSes, the values in each column of a database have a fixed maximal length, which is fixed either implicitly by the datatype, e.g.,\@\xspace{} \(32\) bit for \texttt{INTEGER} columns (in MySQL), or fixed explicitly with the datatype, e.g.,\@\xspace{} \(30\) characters for \texttt{VARCHAR(30)} columns.
For instance, \texttt{ENCODE} converts \say{AB} to the decimal \(3334000000\) for a \texttt{VARCHAR(5)} column.

\begin{algorithm}		
    \fontsize{8}{9.5}\selectfont \begin{algorithmic}[1]
        \State \(l = 0\), \(h = \ensuremath{\abs{eD{}}}\xspace{}\)
        \State \(r = \Call{encode}{\ensuremath{\mathtt{PAE{}\_}\allowbreak{}\mathtt{Dec}(\allowbreak{}\ensuremath{SK_{D}}\xspace{},\allowbreak{}\ensuremath{eD{}_{0}}\xspace)}\xspace}\)
        \State \(N = \Call{encode}{\text{column maximum}}\)  
        \State \(\ensuremath{sVal{}}\xspace{} = (\Call{encode}{\ensuremath{sVal{}}\xspace{}} - r) \% N\)
        \While{\(l < h\)}
            \State \(j = \left \lceil{(l + h) / 2}\right \rceil\)
            \State \(m = \Call{encode}{\ensuremath{\mathtt{PAE{}\_}\allowbreak{}\mathtt{Dec}(\allowbreak{}\ensuremath{SK_{D}}\xspace{},\allowbreak{}\ensuremath{eD{}_{j}}\xspace)}\xspace}\)
            \State \(\ensuremath{cVal{}}\xspace{} = (m - r) \% N\)
            \If{\textcolor{darkestColor}{\((\ensuremath{cVal{}}\xspace{} < \ensuremath{sVal{}}\xspace{})\)} \textcolor{darkColor}{\((\ensuremath{cVal{}}\xspace{} <= \ensuremath{sVal{}}\xspace{})\)}}
                \State \(l = j + 1\)
            \Else
                \State \(h = j\)
            \EndIf
        \EndWhile

        \State \textbf{return} \textcolor{darkestColor}{\((l)\)} \textcolor{darkColor}{\((l - 1)\)}
    \end{algorithmic}
    \caption{\ensuremath{\mathtt{\textcolor{darkestColor}{BinSearchSpecialS}/\textcolor{darkColor}{BinSearchSpecialE}}(\allowbreak{}\ensuremath{eD{}}\xspace{},\allowbreak{}\ensuremath{sVal{}}\xspace{},\allowbreak{}\textit{rndOffset}{},\allowbreak{}\ensuremath{SK_{D}}\xspace{})}\xspace}\label{algo:02:BinarySearchSpecial}
\end{algorithm}

Algorithm~\ref{algo:02:BinarySearchSpecial} first initializes the low and high value of the search.
A value \(r\) is determined by decrypting \ensuremath{eD{}_{0}}\xspace and executing \texttt{ENCODE} on it.
Then, \texttt{ENCODE} is performed on the maximum value that fits the column, which is implicitly defined by the fixed maximal length of the column.
\texttt{ENCODE} is also executed on the search value (\(\ensuremath{sVal{}}\xspace{}\)), \(r\) is subtracted from it and the result is taken modulo \(N\).
All values \(m\) accessed during the search are loaded into the enclave, decrypted and handled as \(\ensuremath{sVal{}}\xspace{}\).
Note that \num{0} is a possible value for \textit{rndOffset}{}, because \textit{rndOffset}{} is chosen uniformly at random between \num{0} and \(\ensuremath{\abs{D{}}}\xspace{} - 1\).
We omit the special handling for brevity. 
Overall, the runtime of \ensuremath{\mathtt{EnclDict}\-\mathtt{Search\_}\allowbreak{}\mathtt{2}}\xspace is logarithmic in \ensuremath{\abs{D{}}}\xspace{} and the encoding introduces only a constant factor compared to \ensuremath{\mathtt{EnclDict}\-\mathtt{Search\_}\allowbreak{}\mathtt{1}}\xspace.

\ensuremath{\mathtt{AttrVect}\-\mathtt{Search\_}\allowbreak{}\mathtt{2}}\xspace linearly scans \ensuremath{AV{}}\xspace{} outside of the enclave and checks if the values \(\ensuremath{v}\xspace{} \in \ensuremath{AV{}}\xspace{}\) fall in either range that was returned by \ensuremath{\mathtt{EnclDict}\-\mathtt{Search\_}\allowbreak{}\mathtt{2}}\xspace.
The RecordID{}s{} (\ensuremath{\mathbf{rid{}}}\xspace{}) of the matching values are returned by this operation.

\textbf{ED3.}
This encrypted dictionary{} combines the repetition option frequency revealing{} and the order option unsorted.
Accordingly, \ensuremath{\mathtt{EncDB\_}\allowbreak{}\mathtt{3}}\xspace performs the frequency revealing{} algorithm and then shuffles the unique values \(\ensuremath{v}\xspace{} \in \ensuremath{D{}}\xspace{}\) randomly resulting in an unsorted dictionary{}.
Afterwards, the ValueID{}s{} in \ensuremath{AV{}}\xspace{} are set such that the split is correct and all values  \(\ensuremath{v}\xspace{} \in \ensuremath{D{}}\xspace{}\) are encrypted with \ensuremath{\mathtt{PAE{}\_}\allowbreak{}\mathtt{Enc}}\xspace{} under \ensuremath{SK_{D}}\xspace{} and a random \ensuremath{IV}\xspace{}.
Figure~\ref{fig:02:exDT1toDT3} (d) shows an example for \ensuremath{\mathtt{EncDB\_}\allowbreak{}\mathtt{3}}\xspace before \ensuremath{\mathtt{PAE{}\_}\allowbreak{}\mathtt{Enc}}\xspace{} is performed.
\ensuremath{\mathtt{EncDB\_}\allowbreak{}\mathtt{3}}\xspace trivially has no order leakage.

ED3's unsorted dictionary{} prevents the use of any search with logarithmic runtime during \ensuremath{\mathtt{EnclDict}\-\mathtt{Search\_}\allowbreak{}\mathtt{3}}\xspace.
Instead, a linear scan over all values \(\ensuremath{c{}}\xspace{} \in \ensuremath{eD{}}\xspace{}\) has to be performed (see Algorithm~\ref{algo:02:dictSearch_7}).
First, \ensuremath{SK_{D}}\xspace{} is derived and used to decrypt the encrypted search range \ensuremath{\tau}\xspace{}.
Then, the algorithm loads each \(\ensuremath{c{}}\xspace{} \in \ensuremath{eD{}}\xspace{}\) into the enclave, decrypts \ensuremath{c{}}\xspace{} and checks if \ensuremath{\mathtt{PAE{}\_}\allowbreak{}\mathtt{Dec}(\allowbreak{}\ensuremath{SK_{D}}\xspace{},\allowbreak{}\ensuremath{c{}}\xspace{})}\xspace falls into \ensuremath{R}\xspace{}.
The result is a list of all matching ValueID{}s{} \ensuremath{\mathbf{vid{}}}\xspace{}.

\begin{algorithm}		
    \fontsize{8}{9.5}\selectfont \begin{algorithmic}[1]
        \State \(\ensuremath{SK_{D}}\xspace{} = \Call{DeriveKey}{\ensuremath{SK_{DB}}\xspace{},colName,tabName}\)
        \State \ensuremath{R}\xspace{} = (\ensuremath{R_s}\xspace{}, \ensuremath{R_e}\xspace{}) = \ensuremath{\big(\ensuremath{\mathtt{PAE{}\_}\allowbreak{}\mathtt{Dec}(\allowbreak{}\ensuremath{SK_{D}}\xspace{},\allowbreak{}\ensuremath{\tau_s}\xspace{})}\xspace,\ensuremath{\mathtt{PAE{}\_}\allowbreak{}\mathtt{Dec}(\allowbreak{}\ensuremath{SK_{D}}\xspace{},\allowbreak{}\ensuremath{\tau_e}\xspace{})}\xspace\big)}
        \State \ensuremath{\mathbf{vid{}}}\xspace{} = \(\emptyset{}\)
        \For{\(i = 0\); \(i < \ensuremath{\abs{D{}}}\xspace{}\); \(i\)++} 
            \State \ensuremath{v}\xspace{} = \ensuremath{\mathtt{PAE{}\_}\allowbreak{}\mathtt{Dec}(\allowbreak{}\ensuremath{SK_{D}}\xspace{},\allowbreak{}\ensuremath{eD{}_{i}}\xspace)}\xspace
            
            \If{\(\ensuremath{R_s}\xspace{} <= \ensuremath{v}\xspace{} <= \ensuremath{R_e}\xspace{}\)} 
                \State \ensuremath{\mathbf{vid{}}}\xspace{}.\Call{append}{i}
            \EndIf
        \EndFor

        \State \textbf{return} \ensuremath{\mathbf{vid{}}}\xspace{}
    \end{algorithmic}
    \caption{\ensuremath{\mathtt{Encl\allowbreak{}Dict\allowbreak{}Search\_}\-\mathtt{7}(\allowbreak{}\ensuremath{\tau}\xspace{},\ensuremath{eD{}}\xspace{})}\xspace\label{algo:02:dictSearch_7}}
\end{algorithm}

\ensuremath{\mathtt{AttrVect}\-\mathtt{Search\_}\allowbreak{}\mathtt{3}}\xspace has to compare every \(\ensuremath{v}\xspace{} \in \ensuremath{AV{}}\xspace{}\) with every \(\ensuremath{u}\xspace{} \in \ensuremath{\mathbf{vid{}}}\xspace{}\) returned by \ensuremath{\mathtt{EnclDict}\-\mathtt{Search\_}\allowbreak{}\mathtt{3}}\xspace.
Thus, the runtime complexity is \(O(\ensuremath{\abs{AV{}}}\xspace{}\cdot\ensuremath{\mathbf{\abs{vid{}}}}\xspace{})\).
However, integers are compared in this case, which is a highly optimized operation in most CPUs\@.
Additionally, \ensuremath{\mathtt{AttrVect}\-\mathtt{Search\_}\allowbreak{}\mathtt{3}}\xspace is easily parallelizable.

\subsubsection*{Frequency smoothing{}}

The main problem of the frequency revealing{} algorithm is that an attacker can learn the frequency of each value \(\ensuremath{D{}_{j}}\xspace \in \ensuremath{D{}}\xspace{}\) even if the values are encrypted.
The reason is that the underlying plaintext values are present only once with a unique ValueID{}.
As a countermeasure, the frequency smoothing{} algorithm bounds the frequency leakage by inserting plaintext duplicates into \ensuremath{D{}}\xspace{} during the column split.
The foundation of this repetition option is the Uniform Random Salt Frequencies method~\cite{Pouliot17}.

In more detail, the frequency smoothing{} algorithm executes a parameterizable and probabilistic random experiment for each unique value \(\ensuremath{v}\xspace{} \in \ensuremath{un(C{})}\xspace{}\) to determine how often \ensuremath{v}\xspace{} should be inserted into \ensuremath{D{}}\xspace{} (see Algorithm~\ref{algo:02:bucketSizes}). 
We say that a plaintext value \ensuremath{v}\xspace{} is split into multiple buckets and every bucket has a specific size.

The number of occurrences of \ensuremath{v}\xspace{} in \ensuremath{C{}}\xspace{} (\ensuremath{\abs{oc(C{},v{})}}\xspace{}) and a maximal bucket size (\ensuremath{bs{}_{\max}}\xspace{}) is passed to this experiment.
The random size for an additional bucket is picked from the discrete uniform distribution \ensuremath{\mathcal{U}\{1, \ensuremath{bs{}_{\max}}\xspace{}\}}\xspace until the total size is above \ensuremath{\abs{oc(C{},v{})}}\xspace{}.
The size of the last bucket is then set such that the total size matches \ensuremath{\abs{oc(C{},v{})}}\xspace{}.
The experiment returns the bucket sizes (\ensuremath{bs_{\text{sizes}}}\xspace{}) and how many buckets were chosen (\ensuremath{\#bs{}}\xspace).
The frequency smoothing{} algorithm inserts \ensuremath{\#bs{}}\xspace{} repetitions of \ensuremath{v}\xspace{} into \ensuremath{D{}}\xspace{}.
For each \(\ensuremath{C{}_{i}}\xspace \in{} \ensuremath{oc(C{},v{}}\xspace){}\), it randomly inserts one of the \ensuremath{\#bs{}}\xspace{} possible ValueID{}s{} into \ensuremath{AV{}_{i}}\xspace.
Each ValueID{} is used exactly as often as defined by \ensuremath{bs_{\text{sizes}}}\xspace{}. 
As a result, the frequency leakage has a bound, because the number of occurrences of each \(ValueID{} \in \ensuremath{AV{}}\xspace{}\) is guaranteed to be between 1 and \ensuremath{bs{}_{\max}}\xspace{}.

\begin{algorithm}[ht]	
    \fontsize{8}{9.5}\selectfont \begin{algorithmic}[1]
        \State \( prevTotal = total = \ensuremath{\#bs{}}\xspace{} = 0 \)
        \State \( \ensuremath{bs_{\text{sizes}}}\xspace{} = \emptyset{}\)

        \While{\( total < \ensuremath{\abs{oc(C{},v{})}}\xspace{}\)}
            \State \( \ensuremath{\#bs{}}\xspace{} \mathrel{+}= 1 \)
            \State \( rnd \xleftarrow{\$} [1, \ensuremath{bs{}_{\max}}\xspace{} ] \)
            \State \( \ensuremath{bs_{\text{sizes}}}\xspace{}.\Call{append}{rnd} \)
            \State \( prevTotal = total \)
            \State \( total \mathrel{+}= rnd \)
        \EndWhile{}
        \State \( \ensuremath{bs_{\text{sizes}}}\xspace{}.\Call{last}{} = \ensuremath{\abs{oc(C{},v{})}}\xspace{} - prevTotal \)
        
        \State \textbf{return} \( \ensuremath{bs_{\text{sizes}}}\xspace{}, \ensuremath{\#bs{}}\xspace{} \)
    \end{algorithmic}
    \caption{\ensuremath{\mathtt{getRndBucketSizes}(\allowbreak{}\ensuremath{\abs{oc(C{},v{})}}\xspace{}, \ensuremath{bs{}_{\max}}\xspace{})}\xspace{}\label{algo:02:bucketSizes}}
\end{algorithm}

\ensuremath{bs{}_{\max}}\xspace{} can be chosen independently for each column.
The selection influences \ensuremath{\abs{D{}}}\xspace{}, which impacts storage efficiency, search time and frequency leakage.
For instance, a large \ensuremath{bs{}_{\max}}\xspace{} leads to few repeating entries in \ensuremath{D{}}\xspace{}, which slightly increases \ensuremath{\abs{D{}}}\xspace{} compared to the frequency revealing{} algorithm. 
This decreases the \ensuremath{\mathtt{EnclDict}\-\mathtt{Search}}\xspace{} performance, because more data needs to be loaded into the enclave, more decryptions are performed, and more comparisons are necessary.
The performance of \ensuremath{\mathtt{AttrVect}\-\mathtt{Search}}\xspace{} also decreases, because more values have to be compared.
A small \ensuremath{bs{}_{\max}}\xspace{} leads to many repetitions in \ensuremath{D{}}\xspace{}, which further increases \ensuremath{\abs{D{}}}\xspace{} and the search time. 
Yet, it leads to a low frequency leakage bound, as each ValueID{} in \ensuremath{AV{}}\xspace{} is present at most \ensuremath{bs{}_{\max}}\xspace{} times.

Next, we explain how the frequency smoothing{} algorithm impacts the three order options, which were introduced in detail before.
We omit the discussion of order leakage as it is independent of the repetition option.

\textbf{ED4.}
\ensuremath{\mathtt{EncDB\_}\allowbreak{}\mathtt{4}}\xspace performs the split of \ensuremath{C{}}\xspace{} according to the frequency smoothing{} algorithm and sorts all values in \ensuremath{D{}}\xspace{} lexicographically determining the order of repetitions randomly. Then, it adjusts the ValueID{}s{} in \ensuremath{AV{}}\xspace{} such that the split is correct while considering how often each ValueID{} can be used, which is defined by \ensuremath{bs_{\text{sizes}}}\xspace{}.
Finally, \ensuremath{\mathtt{EncDB\_}\allowbreak{}\mathtt{4}}\xspace encrypts all \(\ensuremath{v}\xspace{} \in \ensuremath{D{}}\xspace{}\) with \ensuremath{\mathtt{PAE{}\_}\allowbreak{}\mathtt{Enc}}\xspace{} under \ensuremath{SK_{D}}\xspace{} and a random \ensuremath{IV}\xspace{}. 
Note that this only leads to the same ciphertexts with negligible probability, even if the plaintexts are equal.

\ensuremath{\mathtt{EnclDict}\-\mathtt{Search\_}\allowbreak{}\mathtt{4}}\xspace is equal to \ensuremath{\mathtt{EnclDict}\-\mathtt{Search\_}\allowbreak{}\mathtt{1}}\xspace, because leftmost and rightmost binary searches inherently handle repetitions.
The performance penalty compared to ED1 is small, because the binary search only slows down logarithmically with a growing \ensuremath{\abs{D{}}}\xspace{}.
\ensuremath{\mathtt{AttrVect}\-\mathtt{Search\_}\allowbreak{}\mathtt{4}}\xspace equals \ensuremath{\mathtt{AttrVect}\-\mathtt{Search\_}\allowbreak{}\mathtt{1}}\xspace.

\begin{figure}[h]
  \centering
  \includegraphics[width=\linewidth]{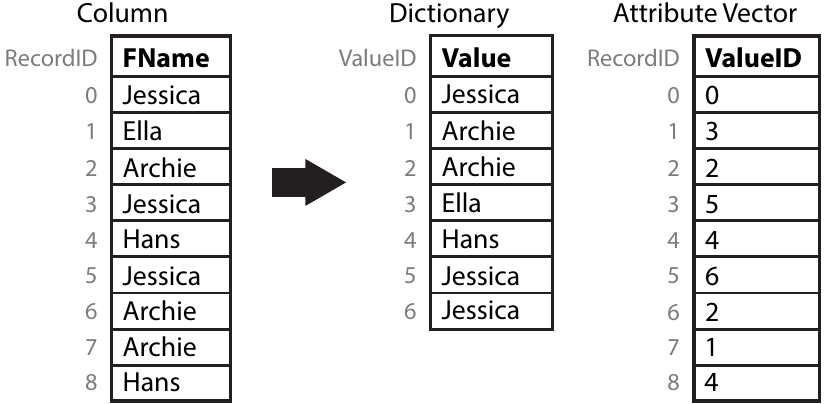}
  \caption{Example for ED5 with \(\ensuremath{bs{}_{\max}}\xspace{} = 3\) and \(\textit{rndOffset}{} = 1\) without encryption}\label{fig:02:exDT5}
\end{figure}

\textbf{ED5.}
For this encrypted dictionary{}, \ensuremath{\mathtt{EncDB\_}\allowbreak{}\mathtt{5}}\xspace performs the split of \ensuremath{C{}}\xspace{} according to the frequency smoothing{} algorithm, rotates the ValueID{}s{} as described in \ensuremath{\mathtt{EncDB\_}\allowbreak{}\mathtt{2}}\xspace, sets the ValueID{}s{} in \ensuremath{AV{}}\xspace{} such that the split is correct (considering \ensuremath{bs_{\text{sizes}}}\xspace{}), and encrypts all \(\ensuremath{v}\xspace{} \in \ensuremath{D{}}\xspace{}\) with \ensuremath{\mathtt{PAE{}\_}\allowbreak{}\mathtt{Enc}}\xspace{} under \ensuremath{SK_{D}}\xspace{} and a random \ensuremath{IV}\xspace{}.
Figure~\ref{fig:02:exDT5} shows an example for ED5 with \(\ensuremath{bs{}_{\max}}\xspace{} = 3\) and \(\textit{rndOffset}{} = 1\) not considering the encryption.

The special binary searches are more complex for ED5 than for ED2, because they have to handle a corner case: the plaintext value of the last and first entry in \ensuremath{D{}}\xspace{} might be equal and present more than two times (as in the example in Figure~\ref{fig:02:exDT5}).
For the same reason, \ensuremath{\mathtt{EnclDict}\-\mathtt{Search\_}\allowbreak{}\mathtt{5}}\xspace has to perform a more complicated postprocessing of \ensuremath{vid{}_{min}}\xspace and \ensuremath{vid{}_{max}}\xspace compared to \ensuremath{\mathtt{EnclDict}\-\mathtt{Search\_}\allowbreak{}\mathtt{2}}\xspace.
The performance penalty compared to ED2 is small, because the binary search slows down logarithmically in \ensuremath{\abs{D{}}}\xspace{}.

\textbf{ED6.}
For columns that are protected with ED6, \ensuremath{\mathtt{EncDB\_}\allowbreak{}\mathtt{6}}\xspace splits \ensuremath{C{}}\xspace{} according to the frequency smoothing{} algorithm, shuffles the values in \ensuremath{D{}}\xspace{}, sets the ValueID{}s{} in \ensuremath{AV{}}\xspace{} such that the split is correct (considering \ensuremath{bs_{\text{sizes}}}\xspace{}), and encrypts all \(\ensuremath{v}\xspace{} \in \ensuremath{D{}}\xspace{}\) with \ensuremath{\mathtt{PAE{}\_}\allowbreak{}\mathtt{Enc}}\xspace{} under \ensuremath{SK_{D}}\xspace{} and a random \ensuremath{IV}\xspace{}.
\ensuremath{\mathtt{EnclDict}\-\mathtt{Search\_}\allowbreak{}\mathtt{6}}\xspace is equal to \ensuremath{\mathtt{EnclDict}\-\mathtt{Search\_}\allowbreak{}\mathtt{3}}\xspace and \ensuremath{\mathtt{AttrVect}\-\mathtt{Search\_}\allowbreak{}\mathtt{6}}\xspace is equal to \ensuremath{\mathtt{AttrVect}\-\mathtt{Search\_}\allowbreak{}\mathtt{3}}\xspace, but frequency smoothing{} se\-verely impacts the performance of these operations.
The reason is that the linear scan of \ensuremath{\mathtt{EnclDict}\-\mathtt{Search\_}\allowbreak{}\mathtt{6}}\xspace needs to load, decrypt, and compare more values.
Additionally, the number of comparisons in \ensuremath{\mathtt{AttrVect}\-\mathtt{Search\_}\allowbreak{}\mathtt{6}}\xspace increases as \ensuremath{\mathtt{EnclDict}\-\mathtt{Search\_}\allowbreak{}\mathtt{6}}\xspace potentially returns more values.

\begin{figure*}[t]
    \centering
    \includegraphics[width=\textwidth]{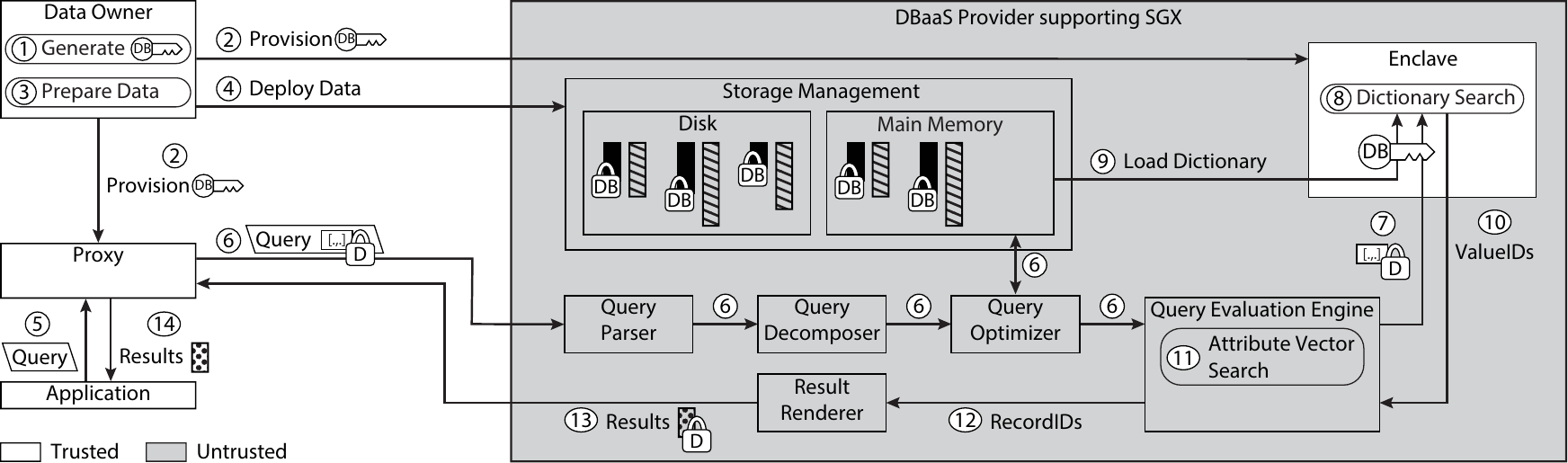}
    \caption{Enc\-DB\-DB{} process flow in detail}
    \label{fig:02:detailDesign}
\end{figure*}

\subsubsection*{Frequency hiding{}}

Now we discuss the frequency hiding{} algorithm, which prevents frequency leakage.
The idea is to add a separate entry into \ensuremath{D{}}\xspace{} for every value in \ensuremath{C{}}\xspace{}, i.e.,\@\xspace{} \(\forall{} i \in [0, \ensuremath{\abs{C{}}}\xspace{} - 1] \colon{} \ensuremath{D{}_{i}}\xspace = \ensuremath{C{}_{i}}\xspace \).
As a result, every \(\ensuremath{v}\xspace{} \in \ensuremath{C{}}\xspace{}\) has \ensuremath{\abs{oc(C{},v{})}}\xspace{} possible ValueID{}s{} at which \ensuremath{v}\xspace{} is stored.
In turn, \(\ensuremath{v}\xspace{} \in \ensuremath{AV{}}\xspace{}\) is set to one of those ValueID{}s{} at random and every ValueID{} is used once.
The resulting dictionary encoding{} does not provide compression anymore (\(\ensuremath{\abs{D{}}}\xspace{} = \ensuremath{\abs{AV{}}}\xspace{}\)), but the frequency of every ValueID{} is perfectly equal, i.e.,\@\xspace{} there is no frequency leakage.

\textbf{ED7, ED8 and ED9.} 
\ensuremath{\mathtt{EncDB\_}\allowbreak{}\mathtt{7}}\xspace, \ensuremath{\mathtt{EncDB\_}\allowbreak{}\mathtt{8}}\xspace and \ensuremath{\mathtt{EncDB\_}\allowbreak{}\mathtt{9}}\xspace execute a split of \ensuremath{C{}}\xspace{} according to the frequency hiding{} algorithm; sort, rotate, and shuffle \ensuremath{D{}}\xspace{}, respectively; adjusts the ValueID{}s{} in \ensuremath{AV{}}\xspace{} such that the split is correct and every index in \ensuremath{D{}}\xspace{} is only used once in \ensuremath{AV{}}\xspace{}; and encrypt all values in \ensuremath{D{}}\xspace{} as described before.

Frequency hiding{} can be interpreted as a special case of frequency smoothing{} with a \ensuremath{bs{}_{\max}}\xspace{} of \num{1}.
Therefore, the \ensuremath{\mathtt{EnclDict}\-\mathtt{Search}}\xspace{} and \ensuremath{\mathtt{AttrVect}\-\mathtt{Search}}\xspace{} operations are equal as described for ED4, ED5, and ED6, and the advantage and disadvantages are equivalent to the ones described for a small \ensuremath{bs{}_{\max}}\xspace{}.

\subsection{EncDBDB in detail}\label{subsec:02:main:designDetail}

In this section, we present how encrypted dictionaries{} are used as foundation for an encrypted DBMS\@.
In one possible Enc\-DB\-DB{} variant, the DBaaS provider is assumed trusted for the initial setup.
The data owner{} can upload plaintext columns and the DBaaS provider can support the data owner{} in choosing a proper encrypted dictionary{} for each column.
Afterwards, the DBaaS performs the appropriate column splits and encryptions.

In the following, however, we discuss another variant in which the DBaaS provider is untrusted also during the setup (see Figure~\ref{fig:02:detailDesign}) and plaintext data never leaves the realm of the data owner{}.
We first describe the system setup, followed by the processing during runtime.

\subsubsection*{Setup}

\circled{1} The data owner{} defines a security parameter \ensuremath{\lambda}\xspace{} and generates a secret key \(\ensuremath{SK_{DB}}\xspace{}=\ensuremath{\mathtt{PAE{}\_}\allowbreak{}\mathtt{Gen}(\allowbreak{}1^\lambda{})}\xspace{}\).

\circled{2} The data owner{} uses SGX's attestation feature to authenticate the DBaaS server's enclave and to establish a secure connection to it (see Section~\ref{subsec:02:background:sgx} for details), which is used to deploy \ensuremath{SK_{DB}}\xspace{} to the enclave.
Additionally, \ensuremath{SK_{DB}}\xspace{} is deployed at the proxy{} via a secure out-of-band mechanism.

\circled{3} The data owner{} takes its plaintext database PDB{} and selects an encrypted dictionary{} for each column \(\ensuremath{C{}}\xspace{} \in PDB{}\).
He performs \ensuremath{\mathtt{EncDB}}\xspace{} as explained in the previous section for each \ensuremath{C{}}\xspace{} according to the selected encrypted dictionary{}.
Each encrypted dictionary{} is encrypted with an individual key \ensuremath{SK_{D}}\xspace{}, which is derived from \ensuremath{SK_{DB}}\xspace{}, the table name, and the column name.
The result is an encrypted database EDB{}{}.

\circled{4} As a last step of the setup, the data owner{} uses the import functionality of the DBaaS provider to deploy EDB{}{}.
The storage management{} of the in-memory database stores all data on disk for persistency and additionally loads it into main memory.

\subsubsection*{Runtime}

From this point on, the application{} can send an arbitrary number of queries, which are processed as follows.

\circled{5} The application{} issues an SQL query \ensuremath{Q{}}\xspace{} to the proxy{}.
W.l.o.g.\@\xspace{} we assume that \ensuremath{Q{}}\xspace{} selects and filters only one column.
The filter can be an equality select, an inequality select, a greater than select (inclusive or exclusive), a less than select (inclusive or exclusive) and a range select (inclusive or exclusive).
The proxy{} converts all filters to a range select with range \(\ensuremath{R}\xspace{} = \ensuremath{\big(\ensuremath{R_s}\xspace{}, \ensuremath{R_e}\xspace{}\big)}\).
For instance, the SQL query \texttt{SELECT FName FROM t1 WHERE FName < \textquotesingle{}Ella\textquotesingle{}} is converted to \texttt{SELECT FName FROM t1 WHERE FName >=} \texttt{\ensuremath{-\ensuremath{\infty}\xspace}\xspace{} and FName < \textquotesingle{}Ella\textquotesingle{}} where \ensuremath{-\ensuremath{\infty}\xspace}\xspace{} is a placeholder for the smallest domain value. 
Next, the proxy{} derives \ensuremath{SK_{D}}\xspace{} using \ensuremath{SK_{DB}}\xspace{}, the table name, and the column name.
Then, it encrypts the range start and end (\ensuremath{R_s}\xspace{} and \ensuremath{R_e}\xspace{}) with \ensuremath{\mathtt{PAE{}\_}\allowbreak{}\mathtt{Enc}}\xspace{} using random initialization vectors.
The resulting encrypted query \ensuremath{eQ{}}\xspace{} of our SQL example is \texttt{SELECT FName FROM t1 WHERE FName >= \ensuremath{\mathtt{PAE{}\_}\allowbreak{}\mathtt{Enc}(\allowbreak{}\ensuremath{SK_{D}}\xspace,\allowbreak{}\ensuremath{IV_{1}}\xspace,\allowbreak{}\ensuremath{-\ensuremath{\infty}\xspace}\xspace{})}\xspace and FName <} \texttt{\ensuremath{\mathtt{PAE{}\_}\allowbreak{}\mathtt{Enc}(\allowbreak{}\ensuremath{SK_{D}}\xspace,\allowbreak{}\ensuremath{IV_{2}}\xspace,\allowbreak{}\texttt{\textquotesingle{}Ella\textquotesingle{}})}\xspace}.
Because of the query conversion, the untrusted DBaaS provider cannot differentiate query types, and due the utilization of probabilistic authenticated encryption{}, it also cannot learn if the values were queried before.
Enc\-DB\-DB{} could also handle other query functionalities, e.g.,\@\xspace{} counts, aggregations, and average calculations, but we do not consider these in this paper, because they are easier to support than range searches.
Other researchers already presented encrypted joins~\cite{arasu2013oblivious,hahn2019,li2008privacy} and it is an interesting future work to support joins while using encrypted dictionaries{}.

\circled{6} \ensuremath{eQ{}}\xspace{} is passed to the query pipeline of the DBaaS provider that is specific to the underlying DBMS\@.
For instance, the query is processed by a \emph{query parser{}}, a \emph{query decomposer{}} and a \emph{query optimizer{}}.
The query optimizer{} selects a query plan and shares it with a \emph{query evaluation engine{}}.
It contains one \ensuremath{\big(\ensuremath{eD{}}\xspace{}, \ensuremath{AV{}}\xspace{}, \ensuremath{\tau}\xspace{}\big)} tuple that is derived from \ensuremath{eQ{}}\xspace{}, i.e.,\@\xspace{} an encrypted dictionary{}, a plaintext attribute vector{} and an encrypted range filter that has to be executed.

\circled{7} The query evaluation engine{} enriches \ensuremath{eD{}}\xspace{} with metadata: the table name, the column name, and the column size.
Then, it passes \ensuremath{\tau}\xspace{} and a reference to \ensuremath{eD{}}\xspace{} to the enclave.

\circled{8} The enclave performs the \ensuremath{\mathtt{EnclDict}\-\mathtt{Search}}\xspace{} operation corresponding to the encrypted dictionary{} of the filtered column.

\circled{9} During this search the necessary dictionary{} entries are loaded from the untrusted realm.

\circled{10} Finally, it returns a list of ValueID{}s{} (\ensuremath{\mathbf{vid{}}}\xspace{}) for which the corresponding values fall into \ensuremath{R}\xspace{}.

\circled{11} The query evaluation engine{} performs the \ensuremath{\mathtt{AttrVect}\-\mathtt{Search}}\xspace{} operation corresponding to the encrypted dictionary{} of the filtered column.
These steps result in a list of RecordID{}s{} (\ensuremath{\mathbf{rid{}}}\xspace{}).

\circled{12} \ensuremath{\mathbf{rid{}}}\xspace{} is passed to a \emph{result renderer{}}, which would use \ensuremath{\mathbf{rid{}}}\xspace{} to prefilter other columns in the same table if a filter query should be executed on them.
Additionally, \ensuremath{\mathbf{rid{}}}\xspace{} would be used if an unconditional select is performed on another column.
One encrypted result column \ensuremath{eC{}}\xspace{} is created by undoing the split in \ensuremath{eD{}}\xspace{} and \ensuremath{AV{}}\xspace{} on all entries in \ensuremath{\mathbf{rid{}}}\xspace{}, i.e.,\@\xspace{} \(\ensuremath{eC{}}\xspace{} = (\ensuremath{eD{}_{j}}\xspace \, | \, j = \ensuremath{AV{}_{i}}\xspace \wedge{} i \in \ensuremath{\mathbf{rid{}}}\xspace{})\).

\circled{13} The result renderer{} enriches \ensuremath{eC{}}\xspace{} with column metadata --- table and column name --- and passed \ensuremath{eC{}}\xspace{} back to the proxy{}.

\circled{14} The proxy{} receives one encrypted column \ensuremath{eC{}}\xspace{} from the DBaaS provider and uses the attached column metadata to derive the column specific key \ensuremath{SK_{D}}\xspace{}.
Every entry in \ensuremath{eC{}}\xspace{} is decrypted individually with \ensuremath{SK_{D}}\xspace{} resulting in one plaintext column \ensuremath{C{}}\xspace{}, which is passed back to the application{}.

Notably, only a very small part of the query processing is done inside the trusted enclave and the required enclave memory is very limited.
There is no need to modify auxiliary database functionalities such as persistency management, multiversion concurrency control or access management.
Still, the complete processing is protected.

\subsection{Dynamic Data}\label{subsec:02:main:dynData}

So far, we only discussed static data, which are prepared by the data owner{} before being uploaded to an Enc\-DB\-DB{}-enabled DBaaS provider.
This is sufficient for most analytical scenarios, because bulk loading of data is often used in this context and complex, read-only queries are executed afterwards~\cite{harizopoulos2006performance,stonebraker2005c}.
For other usage scenarios, we present an approach on how Enc\-DB\-DB{} can support dynamic data, i.e.,\@\xspace{} data insertions, deletions, and updates.

We propose to utilize a concept called \emph{delta store{}} (or differential buffer): the database --- more specifically each column --- is split into a read optimized \emph{main store{}} and a write optimized delta store{} (see~\cite{farber2012sap, hubner2011cost, stonebraker2005c} for more details).
Updates in a column do not change existing rows.
Instead, all data changes are performed in the delta store{}.
New values are simply appended.
Updated values are handled using a validity vector for the two storage concepts.
This vector stores stores a flag for each entry indicating whether or not it is valid.
Deletions are realizable by an update on the validity bit.
The overall state of the column is the combination of both stores.
Thus, a read query becomes more complex: it is executed on both stores normally and then the results are merged while checking the validity of the entries.
The delta store{} should be kept orders of magnitude smaller than the main store{} to efficiently handle read queries.
This is done by periodically merging the data of the delta store{} into the main store{}.
Hübner et al.\ describe different merging strategies~\cite{hubner2011cost}.

For Enc\-DB\-DB{}, any encrypted dictionary{} can be used for the main store{} and ED9 should be employed for the delta store{}.
New entries can simply be appended to a column of type ED9 by reencrypting the incoming value inside the enclave with a random \ensuremath{IV}\xspace{}.
A search in this delta store{} is done by performing the linear scan as defined by \ensuremath{\mathtt{EnclDict}\-\mathtt{Search\_}\allowbreak{}\mathtt{9}}\xspace.
As a result, neither the data order nor the frequency is leaked during the insertion and search.
A drawback of ED9 is that it has a high memory space overhead and low performance.
However, the periodic merges mitigate this problem.
The enclave handles the merging process as follows:
First, it reencrypts every value in \ensuremath{D{}}\xspace{}.
Then, the columns with the rotated order option are randomly rerotated and columns with the unsorted order option are reshuffle.
The process has to be implemented in a way that does not leak the relationship between values in the old and new main store{}, e.g.,\@\xspace{} with oblivious memory primitives~\cite{zeroTrace, zheng2017opaque}. \section{Implementation}\label{sec:02:implementation}

For our experiments we implemented a prototype based on MonetDB, an open-source, column-oriented in-memory DBMS~\cite{boncz_breaking_2008, idreos_monetdb:_2012, boncz2005monetdb}.
MonetDB focuses on read-dominated, analytical workloads and thus perfectly fits our use case.
It is a commercial relational DBMS, which exploits the large main memory of modern computer systems for processing and it uses disk storage for persistency.

MonetDB uses a variant of dictionary encoding{} for all string columns.
The attribute vector contains offsets to the dictionary{}, but the dictionary{} contains data in the order it is inserted (for non-duplicates).
The dictionary{} does not contain duplicates if it is small (below \SI{64}{\kilo\byte}) and a hash table and collision lists are used to locate entries.
The collision list is only used as long as the dictionary{} does not exceed a certain size.
As a result, the dictionary{} might store values multiple times.

The front-end query language of MonetDB is SQL.
We implemented the nine encrypted dictionaries{} as SQL data types in the frontend and new internal data types in the backend.
The encrypted dictionaries{} can be used in SQL create table statements like any other data type, e.g.,\@\xspace{} \texttt{CREATE TABLE t1 (c1 ED7, c2 ED5, \dots{})}.
We further split each dictionary{} into a dictionary head{} and dictionary tail{}.
The dictionary tail{} contains variable length values that are encrypted with AES-128 in GCM mode.
The values are stored sequentially in a random order.
The dictionary head{} contains fixed size offsets to the dictionary tail{} and the values are ordered according to the selected encrypted dictionary{}.
This split is done to support variable length data while enabling an efficient binary search.

For dictionary{} search, we pass a pointer to the encrypted dictionary{} into the enclave and it directly loads the data from the untrusted host process.
Thus, only one context switch is necessary for each query.
Furthermore, all operations mentioned as easily parallelizable run parallel in our implementation.

 \section{Evaluation}\label{sec:02:evaluation}

In this section, we first provide security, storage and performance evaluations of our nine encrypted dictionaries{}.
Based on those, we conclude the section with a usage guideline regarding the different encrypted dictionaries{}.

\subsection{Security Evaluation}\label{subsec:02:evaluation:secEval}

We start this section with a short discussion about enclave code size.
In general, small enclave code size improves the security, as it reduces the probability of security-relevant implementation errors, unintended leakages, and hidden malware.
Our enclave is written in C and besides the Intel SGX SDK (version 2.5), has only 1129 lines of code (LOC).
Only 412 of those LOC are written by us, the remainder are taken up by a big integer library~\cite{bigInteger} used for the dictionary{} search in ED2, ED5 and ED8.
An enclave of this size can be efficiently verified by a user of Enc\-DB\-DB{}.

\begin{figure}[h]
  \centering
  \includegraphics[width=\linewidth]{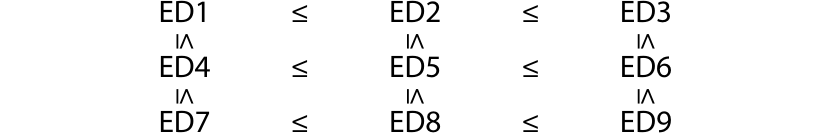}
  \captionof{figure}{Relative security classification. EDX \(\leq\) EDY means that EDY provides the same or better security than EDX.}\label{fig:02:encDTsSecurity}
\end{figure}

Now, we discuss the security of the nine encrypted dictionaries{} under the attacker model defined in Section~\ref{subsec:02:designOverview:attackerModel}, i.e.,\@\xspace{} an honest-but-curious attacker that targets each column independently.
The attacker passively examines the processing of an encrypted dictionary{} \ensuremath{eD{}}\xspace{} and an attribute vector{} \ensuremath{AV{}}\xspace{} in multiple rounds and she knows which encrypted dictionary{} is used.
First, we describe the security of ED1--ED3 and ED7--ED9 by comparing them with security schemes known in literature (see Table~\ref{tab:02:securityClassification}).
A detailed analysis of the different security definitions is beyond the scope of this paper as it is highly data-dependent.
However, we reference known attacks in Table~\ref{tab:02:securityClassification}.
Afterwards, we describe the security of ED4--ED6 relative to the other encrypted dictionaries{}.
The relation between the security provided by the different encrypted dictionaries{} is summarized in Figure~\ref{fig:02:encDTsSecurity}.

\begin{table}[htbp]
  \setlength{\tabcolsep}{2pt}
  \fontsize{9}{10.5}\selectfont \caption{Security of ED1--ED3 and ED7--ED9}\label{tab:02:securityClassification}\centering  
  \begin{tabular}{lllll}
      \toprule
           & \makecell[l]{freq.\\leak.} & \makecell[l]{order\\leak.} & \makecell[l]{comparable\\security} & \makecell[l]{known\\attacks}\\
      \midrule
      ED1   & full    & full    & ideal, determ.~ORE~\cite{boneh2015semantically}  & \cite{TaoInference, grubbs2016leakage}\\ 
      ED2   & full    & bounded & MOPE~\cite{boldyreva_ope_2011}            & \cite{TaoInference, grubbs2016leakage, MOPE_revisited}\\
      ED3   & full    & none    & DET~\cite{bellare2007deterministic}       & \cite{TaoInference, naveed_inference_2015}\\
ED7   & none    & full    & IND-FAOCPA~\cite{kerschbaum2015frequency} & \cite{GrubbsVolume,GuiRangeQueryAttacks,KellarisAttacksOutsourcedDatabases}\\
      ED8   & none    & bounded & IND-CPA-DS~\cite{Kerschbaum17} & \cite{GrubbsVolume,GuiRangeQueryAttacks,KellarisAttacksOutsourcedDatabases}\\
      ED9   & none    & none    & RPE~\cite{lu_privacy-preserving_2012}     & \cite{GrubbsVolume,GuiRangeQueryAttacks,KellarisAttacksOutsourcedDatabases}\\
      \bottomrule
  \end{tabular}\end{table}

ED1 provides protection that is comparable to an ideal, deterministic variant of order-revealing encryption (ORE)~\cite{boneh2015semantically}.
It is an ORE as the order of the values is revealed by a public \say{function} --- the dictionary{}.
It is ideal as neither \ensuremath{eD{}}\xspace{} itself nor the encrypted values leak anything but the order.
It is deterministic as equal plaintexts have the same ciphertext.

The security of modular OPE (MOPE)~\cite{boldyreva_ope_2011} is a lower bound for ED2.
As MOPE, a column protected with ED2 only leaks the \say{modular} order of the values.
MOPE uses deterministic OPE and ED2 uses deterministic ORE, which is more secure.

ED3 has no order leakage, but fully leaks the frequency of all values.
These security features are equivalent to a column protected with deterministic encryption (DET)~\cite{bellare2007deterministic}.

A column protected with ED7 is IND-FAOCPA secure~\cite{kerschbaum2015frequency}.
Each ciphertext is present exactly once in \ensuremath{eD{}}\xspace{} and the assignment of each \ensuremath{AV{}}\xspace{} entry to a ValueID{} is done with the help of a \say{random coinflip} if a plaintext is encrypted multiple times.
Thus, the ValueID{}s{} in \ensuremath{eD{}}\xspace{} form a randomized order (see definition~\cite{kerschbaum2015frequency}) of the plaintext values.

ED8 protects a column with IND-CPA-DS security~\cite{Kerschbaum17}.
\ensuremath{\mathtt{EncDB\_}\allowbreak{}\mathtt{8}}\xspace and the \texttt{Enc} algorithm in~\cite{Kerschbaum17} are different, but the results of the algorithms are equal.
Furthermore, \ensuremath{\mathtt{EnclDict}\-\mathtt{Search\_}\allowbreak{}\mathtt{8}}\xspace matches the \texttt{Search} algorithm in~\cite{Kerschbaum17}.
Therefore, the attacker learns the same information during processing.

The security of a column protected with ED9 is comparable to the security of Range Predicate Encryption (RPE)~\cite{lu_privacy-preserving_2012}.
As defined by RPE's plaintext privacy, \ensuremath{\mathtt{EnclDict}\-\mathtt{Search\_}\allowbreak{}\mathtt{9}}\xspace and \ensuremath{\mathtt{AttrVect}\-\mathtt{Search\_}\allowbreak{}\mathtt{9}}\xspace only leak the information that an entry falls into the search range. 
The \say{predicates} of ED9 are plaintexts encrypted with \ensuremath{\mathtt{PAE{}}}\xspace{} using a random initialization vector, which provides RPE's predicate privacy.

The frequency smoothing{} algorithm used by ED4 makes the ciphertext frequencies close to uniform by randomly selecting a frequency between \(1\) and \ensuremath{bs{}_{\max}}\xspace{}, independent of the plaintext frequency.
As ED1 fully leaks the ciphertext frequency and ED7 hides it completely, the security of ED4 lies between the security of ED1 and ED7.
ED5 is more secure than ED2 and is less secure than ED8 for the same reason.
The same is true for the triple ED6, ED3 and ED9.
The frequency smoothing{} algorithm is based on an algorithm described in~\cite{Pouliot17} and the authors only state that the last frequency is not selected from the same distribution, which might give an advantage to an attacker. 
An in-depth security evaluation is an open research question.

\subsection{Storage Evaluation}\label{subsec:02:evaluation:storageEval}

For our storage evaluation, we use a snapshot of a real-world SAP customer's business warehouse (BW) system.
The largest columns contain \(168.7\) million data values.
To evaluate the influence of the number of unique values to our algorithms, we search for columns having the same size, but different distributions.
The dataset contains 30 large columns with \(10.9\) million values.
We present the results for two extreme cases: C1 with 6.96 million unique values and C2 with \num{13361}.

Table~\ref{tab:02:fileSizes} presents the storage space requirements of different variants.
The plaintext file contains all plaintext values present in the column without any compression.
This file is comparable to a plaintext column for which dictionary encoding{} is not used.
The encrypted file contains every value from the plaintext file, but individually encrypted with \ensuremath{\mathtt{PAE{}}}\xspace{}, which has the same storage requirements as an encrypted column without dictionary encoding{}.
MonetDB's storage requirements are presented as a baseline.

\begingroup
\setlength{\tabcolsep}{4pt}
\begin{table}[htbp]
  \fontsize{9}{10.5}\selectfont \caption{Storage size of various variants}\label{tab:02:fileSizes}\centering
  \begin{tabular}{lrrr}
      \toprule
        & size C1 & size C2\\
      \midrule
      Plaintext file                                              & \SI{136}{\mega\byte}   & \SI{93}{\mega\byte} \\
      Encrypted file                                              & \SI{437}{\mega\byte}   & \SI{392}{\mega\byte} \\
      MonetDB                                                     & \SI{132}{\mega\byte}   & \SI{43}{\mega\byte}  \\
      ED1/ED2/ED3                      & \SI{347}{\mega\byte}   & \SI{22}{\mega\byte}  \\
      ED4/ED5/ED6, \(\ensuremath{bs{}_{\max}}\xspace{} = 100\)  & \SI{347}{\mega\byte}   & \SI{56}{\mega\byte}  \\
      ED4/ED5/ED6, \(\ensuremath{bs{}_{\max}}\xspace{} = 10\)   & \SI{367}{\mega\byte}   & \SI{123}{\mega\byte} \\
      ED4/ED5/ED6, \(\ensuremath{bs{}_{\max}}\xspace{} = 2\)    & \SI{455}{\mega\byte}   & \SI{331}{\mega\byte} \\        
      ED7/ED8/ED9                      & \SI{515}{\mega\byte}   & \SI{475}{\mega\byte} \\
      \bottomrule
    \end{tabular}\end{table}
\endgroup

The size of the plaintext files decreases from C1 to C2, because the strings in these columns are 12 and 10 characters long. As expected, we see that Enc\-DB\-DB{} requires less space if fewer unique values are present.
We see that for C2 protected with ED1, ED2, or ED3, Enc\-DB\-DB{} requires less storage space than the plaintext file, i.e.,\@\xspace{} less space than a plaintext column without dictionary encoding{}.
We also see a further expected behavior: a smaller \ensuremath{bs{}_{\max}}\xspace{} increases the required storage space as more duplicates are stored.

Note that the encrypted dictionaries{} are stored outside of the enclave and individual values are loaded and decrypted.
Hence, the restricted enclave space does not constitute a limitation for Enc\-DB\-DB{}.

\subsection{Performance Evaluation}\label{subsec:02:evaluation:perfEval}

For the performance evaluation, we use the same columns introduced in the storage evaluation.
Besides the original columns, which we call \emph{full datasets}, we sample datasets from \(1\) to \(10\) million records using the distribution and values of the original columns.

MonetDB is used as one baseline measurement in our experiments to compare ourselves against a commercial plaintext DBMS\@.
Additionally, we implement \emph{Plain\-DB\-DB{}} --- a plaintext variant of Enc\-DB\-DB{}.
Plain\-DB\-DB{} uses the same algorithms as Enc\-DB\-DB{}, but the dictionaries{} are plaintext and the algorithms are processed without an enclave.
We use Plain\-DB\-DB{} as a second baseline to evaluate the performance overhead of encryption and SGX\@.

All experiments are performed with the confidential computing offering of Microsoft Azure.
We use a DC4s machine with \SI{16}{\gibi\byte} RAM and 4 vCPU cores of an Intel Xeon E-2176G CPU @ 3.70GHz.
All presented latencies measure the processing time spent at the server excluding any network delay or processing at the proxy or client.
Our protocol runs in one round and only encrypts the values in the query.
Thus, the communication and latency overhead compared to any database in the cloud is negligible.

\begin{figure}[h]
  \centering
  \includegraphics[]{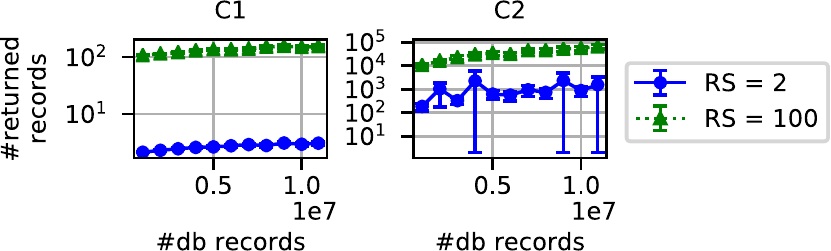}
  \caption{Average number results returned by 500 random range queries for columns C1 and C2 (95\% confidence interval; note that logarithmic y-axis distorts error bars)}\label{fig:02:returnedTuples}
\end{figure}

We use the term \emph{range size} (\ensuremath{RS}\xspace{}) to describe how many consecutive unique values from the dataset are searched in a range query, i.e.,\@\xspace{} if \(sorted(\ensuremath{un(C{})}\xspace{}) = (\allowbreak{}\ensuremath{v_{0}}\xspace,\allowbreak{}\ldots{},\allowbreak{}\ensuremath{v_{\ensuremath{\abs{un(C{})}}\xspace - 1}}\xspace ) \) is a sorted list of all unique values in \ensuremath{C{}}\xspace{}, then \ensuremath{RS}\xspace{} defines the search range \(\ensuremath{R}\xspace{} = [\ensuremath{v_{i}}\xspace, \ensuremath{v_{i + \ensuremath{RS}\xspace{} - 1}}\xspace]\) for \(i \in [0, \ensuremath{\abs{un(C{})}}\xspace{} - \ensuremath{RS}\xspace{}]\).
For every dataset and encrypted dictionary{}, we perform 500 random range queries with range sizes 2 and 100.
The same random range queries are executed for MonetDB, Plain\-DB\-DB{}, and Enc\-DB\-DB{}.
Note that the number of result rows returned by the server is greater than \ensuremath{RS}\xspace{} if a value in the search range is present multiple times in the column (see Figure~\ref{fig:02:returnedTuples}).
For instance, \num{65067} values are returned on average for the full dataset of C2 and \(\ensuremath{RS}\xspace{} = 100\).

\begin{figure*}[h!]
  \centering
  \includegraphics[width=\linewidth]{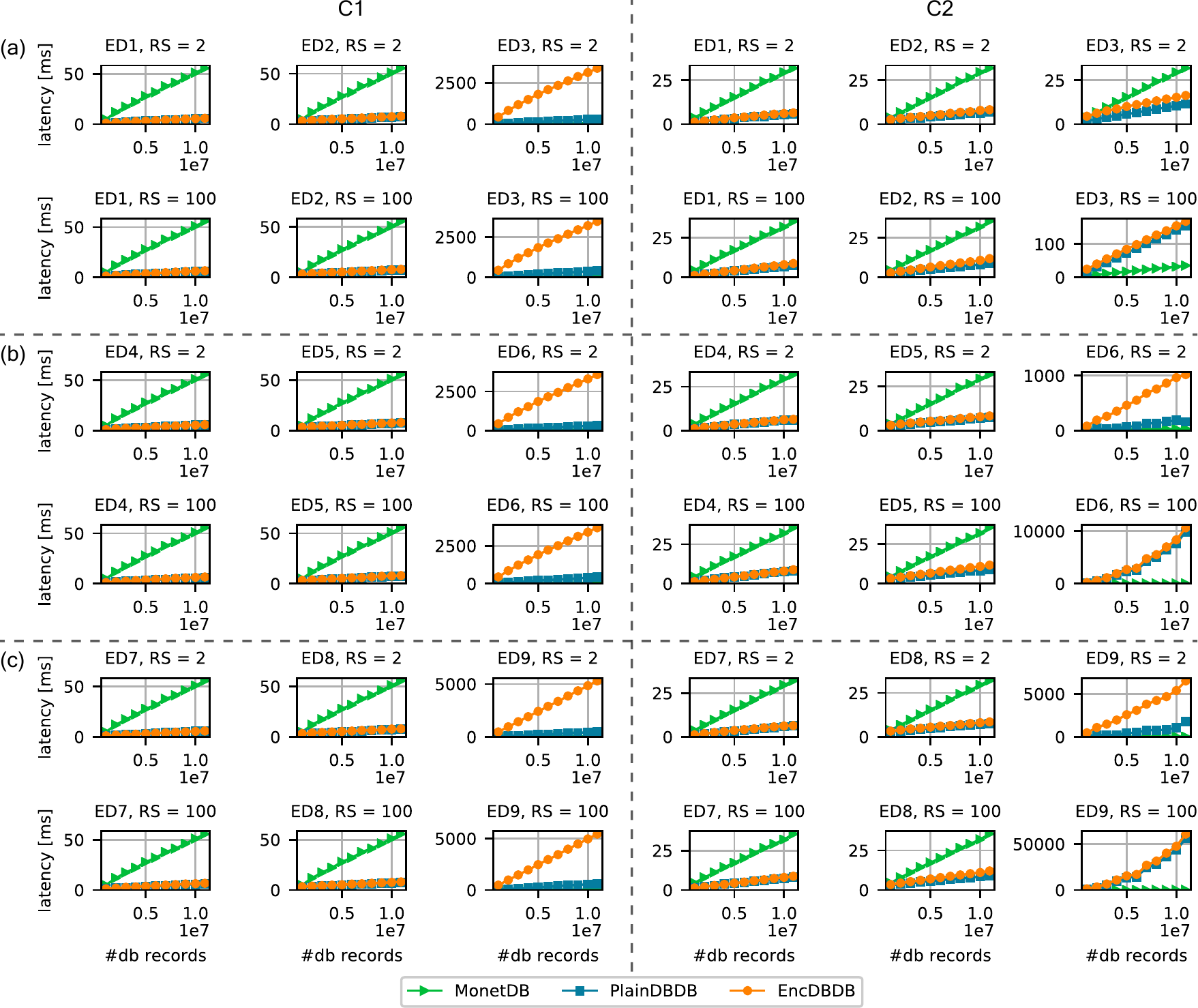}
  \caption{Average latencies of 500 random range queries for columns C1 and C2, which are protected by (a) ED1--ED3, (b) ED4--ED6, and (c) ED7--ED9 (95\% confidence interval not big enough to be visible)}\label{fig:02:perfEval}
\end{figure*}

\textbf{ED1.}
The first and fourth column in Figure~\ref{fig:02:perfEval} (a) present the latencies of ED1 for C1 and C2 and the range sizes 2 and 100.
We highlight three observations from these plots.
First, Enc\-DB\-DB{} and Plain\-DB\-DB{} outperform MonetDB for both range sizes at both columns.
The main reason is that MonetDB's attribute vector search performs a linear number of string comparisons.
In contrast, Enc\-DB\-DB{} and Plain\-DB\-DB{} require only a logarithmic number of string comparisons in the dictionary{} search and a linear number of integer comparisons in the attribute vector{} search.
Second, Enc\-DB\-DB{} slows down if a column with equal size has less unique values: the average latencies increase from \SI{6.55}{\milli\second} at C1 to \SI{8.79}{\milli\second} at C2 for the full dataset and \(\ensuremath{RS}\xspace{} = 100\).
This seems counterintuitive, because fewer unique values result in a smaller dictionary{} size (\ensuremath{\abs{D{}}}\xspace{}) improving the dictionary{} search performance.
However, only logarithmically fewer decryptions and string comparisons are necessary in the dictionary{} search, but many results are returned by the attribute vector{} search (see Figure~\ref{fig:02:returnedTuples}).
As a result, the DBMS has to spend more time for tuple-reconstruction, i.e.,\@\xspace{} to build the result set based on the found RecordID{}s{} and the dictionary{}.
Third, encryption is cheap: the average latency overhead of Enc\-DB\-DB{} compared to Plain\-DB\-DB{} is \SI{0.36}{\milli\second} (\SI{8.9}{\percent}).
The overhead is minor for two reasons: (1) as explained in the implementation section, we only requires one context switch per column, which is negligible in the overall latency and (2) we only use hardware-supported AES-GCM encryption.

\textbf{ED2.}
The second and fifth column in Figure~\ref{fig:02:perfEval} (a) present the latencies of ED2.
The main observation is that the latency of Enc\-DB\-DB{} and Plain\-DB\-DB{} is almost equal to the latency of ED1 for the two columns.
The only difference between ED1 and ED2 is that ED2 uses a special binary search and post-processing of the resulting ValueID{}s{} to handle the random rotation, which introduces only a minor overhead.
In fact, the average latency overhead from ED1 to ED2 is \SI{1.88}{\milli\second} for Enc\-DB\-DB{}.

\textbf{ED3.}
The third and sixth column in Figure~\ref{fig:02:perfEval} (a) show the latencies of ED3.
We observe that the average latencies of Plain\-DB\-DB{} and Enc\-DB\-DB{}, and their relative latency differences, severely depend on the number of unique values and the range size (\ensuremath{RS}\xspace{}).
C2 has a smaller \ensuremath{\abs{D{}}}\xspace{} than C1, which decreases the latency of the linear dictionary{} search and therefore the average latency of the query execution.
Additionally, a smaller \ensuremath{\abs{D{}}}\xspace{} decreases the number of necessary decryptions for Enc\-DB\-DB{} and therefore the relative latency difference between Plain\-DB\-DB{} and Enc\-DB\-DB{}.

\textbf{ED4, ED5, ED6.}
Figure~\ref{fig:02:perfEval} (b) presents the latency plots for ED4--ED6.
The latencies of MonetDB obviously do not change. In the following, we focus on Enc\-DB\-DB{} discussing the latencies for ED4--ED6 compared to ED1--ED3.
\ensuremath{\abs{D{}}}\xspace{} is larger for ED4--ED6, because the frequency smoothing{} algorithm adds duplicates to \ensuremath{D{}}\xspace{} (\(\ensuremath{bs{}_{\max}}\xspace{} = 10\) in our experiments).
For ED4 and ED5, \ensuremath{\abs{D{}}}\xspace{} influences the latency only logarithmically.
Compared to ED1 and ED2, the average overheads are only \SI{0.002}{\milli\second} and \SI{0.11}{\milli\second}, respectively.
At ED6, the dictionary{} search might return more than \(x\) ValueID{}s{} for the range size \(x\) as \ensuremath{eD{}}\xspace{} contains duplicate plaintexts.
Every returned value has to be compared to each attribute vector{} entry.
This increases the average latencies for the full dataset at \(\ensuremath{RS}\xspace{} = 100\) to \SI{3.59}{\second} and \SI{10.64}{\second} for C1 and C2.

\textbf{ED7, ED8, ED9.}
Figure~\ref{fig:02:perfEval} (c) presents the latency plots for ED7--ED9.
We again focus on Enc\-DB\-DB{}'s latency in ED7--ED9 compared to ED1--ED3.
Compared to ED1 and ED2, the average overheads of ED7 and ED8 are \SI{0.01}{\milli\second} and \SI{0.23}{\milli\second}, respectively.
For the full dataset at \(\ensuremath{RS}\xspace{} = 100\), the average latencies of ED9 increase to \SI{5.43}{\second} and \SI{60.82}{\second} for C1 and C2, respectively.

\subsection{Usage Guideline}\label{subsec:02:evaluation:usageGuidelines}

According to the security sensitivity of the data owner{}, an encrypted dictionary{} can be select per column.
If plaintext is not an option, but the weakest security level is acceptable, ED1 can be used.
It has a small storage size and it is almost as fast as Plain\-DB\-DB{}, even with different range sizes and unique value amounts.
If order leakage should be reduced and a minor performance overhead is acceptable, ED2 is preferable over ED1.
If order leakage is not acceptable, a column contains few unique values, and \ensuremath{RS}\xspace{} is small, ED3 has a practical overhead.
For instance, Enc\-DB\-DB{}'s average latency overhead from ED1 to ED3 for C2 and \(\ensuremath{RS}\xspace{} = 2\) is \SI{6.87}{\milli\second}.
If the frequency leakage should be bounded, ED5 can be used with a minor performance and storage overhead compared to ED2.
In many cases, ED5 is the best security, latency and storage tradeoff among our encrypted dictionaries{}.
If security and latency are critical, but not storage size, ED8 is the most favorable encrypted dictionary{}.
If security is the main objective of a column, ED9 should be used. \section{Related Work}\label{sec:02:relatedWork}

In this section, we compare Enc\-DB\-DB{} to TEE-based encrypted databases, software-only encrypted databases, and searchable encryption.

\subsection{TEE-based Encrypted Databases}\label{subsec:02:relatedWork:teeBased}
In the following, we outline TEE-based approaches ranging from large to small enclave sizes, and classify Enc\-DB\-DB{} accordingly.

Haven~\cite{baumann_shielding_2014} and SCONE~\cite{scone} are approaches to shield complete applications on an untrusted system using SGX\@.
Unmodified applications should be executable inside an SGX enclave, which could also be an off-the-shelf DBMS\@.
However, a complete DBMS with millions of lines of code is prone to security-relevant implementation errors or side-channel leakages that could leak arbitrary data from the enclave.
Furthermore, no TEE on the market does support the huge enclaves that are necessary for this concept.

Priebe et al.\ proposed EnclaveDB~\cite{EnclaveDB}, a protected database engine that uses a TEE to provide confidentiality, integrity, and freshness for OLTP workloads.
EnclaveDB has a large TCB, as the tables, indices, metadata, query engine, transaction manager and stored procedures are loaded into the TEE\@.
The problems described for Haven and SCONE are only slightly less severe.
Especially, it still does not fit into an existing TEE\@.
Furthermore, all possible queries have to be known in advance.

ObliDB~\cite{ObliDB18} is an SGX based encrypted database that hides the access pattern using oblivious query processing algorithms on a $B^+$-tree\xspace{} index or a linear array.
The additional protection introduces a latency overhead of 240\% compared to a plaintext database.
Additionally, ObliDB lacks transaction management and disk persistency.

HardIDX~\cite{fuhry2017hardidx} uses SGX to protect one specific data structure, a $B^+$-tree\xspace{}.
Equality and range searches are done inside the enclave and either the whole dataset at once or parts on demand are loaded into enclave memory.
Only a few megabytes of enclave memory are necessary, and the enclave has only a few lines of code.
However, a $B^+$-tree\xspace{} is only presented as a building block of an encrypted database.

TrustedDB~\cite{bajaj2014trusteddb}, Cipherbase~\cite{transactioCipherbase}, and StealthDB~\cite{gribov2017stealthdb} use a secure co-processor, an FPGA, and SGX as TEE, respectively.
They have the smallest enclave size by putting the execution of individual operators, e.g.,\@\xspace{} \(<\), \(>\), and \(=\) into a TEE\@.
The operations are executed on encrypted data and the results are passed back.
Only minor changes to an application (e.g.,\@\xspace{} a database) are necessary as plaintext operations are just replaced by protected operators.
However, much information is leaked as an attacker learns the result of each operation.
Only the processing of data structures, e.g.,\@\xspace{} dictionaries{} or $B^+$-trees\xspace{}, is protected, but the authors do not consider the inherently leaked information about the relation of individual data values. 

Enc\-DB\-DB{} follows the same design philosophy as HardIDX\@: keep the enclave code and the required enclave memory as small as possible without leaking every individual decision by processing a data structure inside an enclave.
As a main difference, we integrate Enc\-DB\-DB{} into a DBMS\@.

\subsection{Software-Only Encrypted Databases}\label{subsec:02:relatedWork:encDB}

Software only encrypted databases, such as CryptDB~\cite{popa_cryptdb:_2011} and Monomi~\cite{Monomi}, use property-preserving encryption for efficient search.
Every database functionality requires its own encryption scheme with additional storage overhead.
For instance, deterministic encryption~\cite{bellare2007deterministic} is used to support equality selects, and OPE~\cite{agrawal_order_2004,boldyreva_ope_2009,boldyreva_ope_2011,kerschbaum_optimal_2014} allows range queries.
Naveed et al.~\cite{naveed_inference_2015} presented practical ciphertext-only attacks on property-preserving encryption and further attacks followed~\cite{durak2016else,grubbs2016leakage}.
In Enc\-DB\-DB{}, equality and range queries are handled by one encryption scheme with a small performance and storage overhead.
Some encrypted dictionaries{} of Enc\-DB\-DB{} are affected by these attacks, but the data owner{} can freely choose a security level that fits his requirements and all functionality is still supported.

Other approaches for a secure DBMS allowing range query evaluation have been published:
Cash et al.~\cite{cash2013highly} introduce a protocol that allows evaluation of boolean queries on encrypted data.
Faber et al.~\cite{faber2015rich} extend this protocol to support range queries but either leak additional information on the queried range or the result set contains false positives.  
Pappas et al.~\cite{pappas2014blind} evaluate encrypted bloom filters using secure multiparty computation.
However, in order to achieve practical efficiency, they propose to split the server into two non-colluding parties. 
Egorov et al.~\cite{egorov_zerodb_2016} presented ZeroDB, a database that enables a client to perform equality and range searches with the help of $B^+$-trees\xspace.
It uses an interactive protocol requiring many rounds and thus is not usable for network-sensitive cloud computing.
Enc\-DB\-DB{} does neither require an additional party nor multiple rounds.

\subsection{Searchable Encryption}\label{subsec:02:relatedWork:searchEnc}

Song et al.\ introduced the first searchable encryption schemes for single plaintexts~\cite{song_practical_2000}.
In order to improve performance, Goh~\cite{goh_secure_2003} and Curtmola et al.~\cite{curtmola_searchable_2006} introduced encrypted (inverted) indices.
However, these encryption schemes can only search for keyword equality and not ranges.

The first range-searchable scheme by Boneh and Waters encrypts every entry linear in the size of the plaintext domain~\cite{BW07}.
The first scheme with logarithmic storage size per entry in the domain was proposed by Shi et al.\ in~\cite{shi_multi-dimensional_2007}.
Their security model (match-revealing) is somewhat weaker than standard searchable encryption.
The construction is based on inner-product predicate encryption which has been made fully secure by Shen et al.\ in~\cite{shen_predicate_2009}.
All of these schemes have linear search time.
Lu built the range-searchable encryption from~\cite{shen_predicate_2009} into an index in~\cite{lu_privacy-preserving_2012}, thereby enabling polylogarithmic search time.
Hahn and Kerschbaum proposed a construction building the search index incrementally~\cite{hahn2016poly}. 
By doing so, they still support amortized polylogarithmic search time but increase security properties for non-queried values.
Demertzis et al.\ presented multiple constructions that improve the constant factor of a range search~\cite{demertzis_2016}.
However, their construction without prohibitive storage cost and false positives (Logarithmic-URC) requires already more than a second to perform a range search within \num{100000} values~\cite{fuhry2017hardidx}.
Enc\-DB\-DB{} operates on millions of entries in milliseconds.

 \section{Conclusion}\label{sec:02:conclusion}

In this paper, we introduced Enc\-DB\-DB{} --- a high-perfor\-mance, encrypted cloud database supporting analytic queries on large datasets.
Enc\-DB\-DB{} provides nine different encrypted dictionaries{} with distinct security, performance and storage efficiency tradeoffs.
Even with no frequency leakage and bounded order leakage, range queries on datasets with millions of encrypted entries are executed within milliseconds.
If some frequency leakage is acceptable, the compressed encrypted data requires less space than a plaintext column.
Moreover, the TCB of Enc\-DB\-DB{} consists only of 1129 lines of code exposing only a small attack surface.
With those features, Enc\-DB\-DB{} is ideally suited for an entity that wants to outsource its sensitive data to an untrusted cloud environment. 
\bibliographystyle{abbrv}
\bibliography{ms}      

\end{document}